\documentclass[11pt]{article}
\usepackage{amssymb}

\usepackage{graphicx}
\usepackage{amsmath}
\usepackage{makeidx}
\usepackage{indentfirst}

\newcounter{resultnum}[section]

\newcounter{conclusionnum}[section]

\newcounter{conditionnum}[section]

\newcounter{conjecturenum}[section]

\newcounter{examplenum}[section]

\newcounter{exercisenum}[section]

\newcounter{lemmanum}[section]

\newcounter{notationnum}[section]

\newcounter{theoremnum}[section]

\newcounter{definitionnum}[section]

\newcounter{corollarynum}[section]

\newcounter{remarknum}[section]

\newcounter{propositionnum}[section]

\newcounter{acknowledgementnum}[section]

\newcounter{algorithmnum}[section]

\newcounter{axiomnum}[section]

\newcounter{casenum}[section]

\newcounter{claimnum}[section]

\newcounter{summarynum}[section]

\newcounter{problemnum}[section]

\begin{document}

\title{Branes and Quantization for an A--Model\\
Complexification of Einstein Gravity\\
in Almost K\"{a}hler Variables}
\date{April 1, 2009}
\author{ Sergiu I. Vacaru \thanks{
the affiliation for Fields Institute is for a former visiting position;
\newline
Sergiu.Vacaru@gmail.com;\ http://www.scribd.com/people/view/1455460-sergiu }
\\
{\quad} \\
{\small {\textsl{Faculty of Mathematics, University "Al. I. Cuza" Ia\c si},}
}\\
{\small {\textsl{\ 700506, Ia\c si, Romania}} }\\
{\small and}\\
{\small \textsl{The Fields Institute for Research in Mathematical Science}}
\\
{\small \textsl{222 College Street, 2d Floor, Toronto \ M5T 3J1, Canada}} }
\maketitle

\begin{abstract}
The general relativity theory is redefined equivalently in almost K\"{a}hler
variables: symplectic form, $\mathbf{\theta}[\mathbf{g}],$ and canonical
symplectic connection, $\widehat{\mathbf{D}}[\mathbf{g}]$ (distorted from
the Levi--Civita connection by a tensor constructed only from metric
coefficients and their derivatives). The fundamental geometric and physical
objects are uniquely determined in metric compatible form by a (pseudo)
Riemannian metric $\mathbf{g}$ on a manifold $\mathbf{V}$ enabled with a
necessary type nonholonomic $2+2$ distribution. Such nonholonomic symplectic
variables allow us to formulate the problem of quantizing Einstein gravity
in terms of the A--model complexification of almost complex structures on $%
\mathbf{V},$ generalizing the Gukov--Witten method \cite{gukwit}. Quantizing
$(\mathbf{V},\mathbf{\theta}[\mathbf{g}],\ \widehat{\mathbf{D}}[\mathbf{g}%
]), $ we derive a Hilbert space as a space of strings with two A--branes
which for the Einstein gravity theory are nonholonomic because of induced
nonlinear connection structures. Finally, we speculate on relation of such a
method of quantization to curve flows and solitonic hierarchies defined by
Einstein metrics on (pseudo) Riemannian spacetimes.

\vskip0.3cm

\textbf{Keywords:}\ quantum gravity, Einstein gravity, nonholonomic
manifolds, symplectic variables, nonlinear connections, strings and A--branes

\vskip3pt 2000 MSC:\ 83C45, 81S10, 53D55, 53B40, 53B35, 53D50

PACS:\ 04.20.-q, 02.40.-k, 02.90.+g, 02.40.Yy
\end{abstract}

\tableofcontents

\section{Introduction}

In this paper we address the question of quantization of an A--model
complexification of spacetime in general relativity following a new
perspective on symplectic geometry, branes and strings proposed in Ref. \cite%
{gukwit} (the Gukov--Witten approach). Symplectic techniques has a long
history in physics (see, for instance, \cite{guillst}) and, last two
decades, has gained more and more interest in the theory of deformation
quantization \cite{fedos,konts,karabeg1,castro1,castro2}. Recently, an
approach based on Fedosov quantization of Einstein gravity \cite{vpla} and
Lagrange--Finsler and Hamilton--Cartan spaces \cite{vfqlf,avqg5} was
elaborated in terms of almost K\" ahler variables on nonholonomic manifolds
(see \cite{vsgg,vrflg}, for a review of methods applied to standard theories
of physics, and \cite{ma2,avqg5,bejf}, for alternative geometrizations of
nonholonomic mechanics on manifolds and/or bundle spaces).\footnote{%
A pair $(\mathbf{V},\mathcal{N})$, where $\mathbf{V}$ is a manifold and $%
\mathcal{N}$ is a nonintegrable distribution on $\mathbf{V}$, is called a
nonholonomic manifold. We emphasize that in this paper we shall not work
with classical physical theories on (co) tangent bundles but only apply in
classical and quantum gravity certain methods formally elaborated in the
geometry of Lagrange--Finsler spaces and nonholonomic manifolds. Readers may
find in Appendix and Section \ref{sakv} the most important definitions and
formulas from the geometry of nonholonomic manifolds and a subclass of such
spaces defined by nonlinear connection, N--connection, structure (i.e.
N--anholonomic manifolds).}

The problem of quantization of nonlinear physical theories involves a grate
amount of ambiguity because the quantum world requests a more refined and
sophisticate description of physical systems than the classical approach.
Different types of quantization result in vary different mathematical
constructions which lead to inequivalent quantizations of the same classical
theories (the most important approaches to quantization are discussed in %
\cite{gukwit,alienglis,vloopdq}, see also references therein).

Deformation quantization was concluded to be a systematic mathematical
procedure but considered that it is not a quantization in a standard manner %
\cite{gukwit}. This is because a deformation quantization of the ring of
holomorphic functions on a complex symplectic/Poisson manifold requires not
arbitrary choices but quantization does. It does not use deformations with a
complex parameter, works with deformations over rings of formal power series
and does not lead to a natural Hilbert space on which the deformed algebra
acts. Nevertheless, different methods and results obtained in deformation
quantization play a very important role in all approaches to quantization,
as a consequent geometric formalism in nonlinear functional analysis.
Perhaps, the most important results in deformation quantization can be
redefined in the language of other approaches to quantization which is very
useful for developing new methods of quantization.

In this paper we consider a new perspective on quantization of Einstein
gravity based on two--dimensional sigma--models, following the A--model
quantization via branes proposed in \cite{gukwit}. Our purpose is to get
closer to a systematic theory of quantum gravity in symplectic variables
related to an almost K\"{a}hler formulation of general relativity. The novel
results in this paper are those that we propose an explicit application of
the Gukov--Witten quantization method to gravity and construct a Hilbert
space for Einstein spaces parametrizing it as the spaces of two nonholonomic
A--brane strings. We relate the constructions to group symmetries of curve
flows, bi--Hamilon structures and solitonic hierarchies defined by (pseudo)
Riemannian/ Einstein metrics.

The new results have a strong relationship to our former results on
nonholonomic Fedosov quantization of gravity \cite{vpla} and
Lagrange--Finser/ Hamilton--Cartan systems \cite{vfqlf,avqg5}.
Geometrically, such relations follow from the fact that in all cases the
deformation quantization of a nonholonomic complex manifold, constructed
following the Gukov--Witten approach, produces a so--called distinguished
algebra (adapted to a nonlinear connection structure) that then acts in the
quantization of a real almost K\"{a}hler manifold. It is obvious that
different attempts and procedures to quantize gravity theories are not
equivalent. For such generic nonlinear quantum models, it is possible only
to investigate the conditions when the variables from one approach can be
re--defined into variables for another one. Then, a more detailed analysis
allows us to state the conditions when physical results for one quantization
are equivalent to certain ones for another quantum formalism.

The paper is organized as follows: In Section 2, we provide an introduction
in the almost K\"{a}hler model of Einstein gravity. Section 3 is devoted to
formulation of quantization method for the A--model with nonholonomic
branes. In section 4, an approach to Gukov--Witten quantization of the
almost K\"{a}hler model of Einstein gravity is developed. Finally, we
present conclusions in section 5. In Appendix, we summarize some important
component formulas necessary for the almost K\"{a}hler formulation of
gravity.

Readers may consult additionally the Refs. \cite%
{vexsol,vegnakglfdq,vrflg,vsgg} on conventions for our system of denotations
and reviews of the geometric formalism for nonholonomic manifolds, and
various applications in standard theories of physics.

\section{Almost K\"{a}hler Variables in Gravity}

\label{sakv}The standard formulation of the Einstein gravity theory is in
variables $(\mathbf{g},\ ^{\mathbf{g}}\nabla ),$ for $\ ^{\mathbf{g}}\nabla
=\nabla \lbrack \mathbf{g}]=\{\ _{\shortmid }^{\mathbf{g}}\Gamma _{\ \beta
\gamma }^{\alpha }=\ _{\shortmid }\Gamma _{\ \beta \gamma }^{\alpha }[%
\mathbf{g}]\}$ being the Levi--Cevita connection completely defined by a
metric $\mathbf{g=\{\mathbf{g}_{\mu \nu }\}}$ on a spacetime manifold $%
\mathbf{V}$ and constrained to satisfy the conditions $\ ^{\mathbf{g}}\nabla
\mathbf{g}=0$ and $\ _{\shortmid }^{\mathbf{g}}T_{\ \beta \gamma }^{\alpha
}=0,$ where $\ _{\shortmid }^{\mathbf{g}}T$ is the torsion of $\ ^{\mathbf{g}%
}\nabla .$\footnote{%
We follow our conventions from \cite{vrflg,vsgg} when 'boldfaced'' symbols
are used for spaces and geometric objects enabled with (or adapted to) a
nonholonomic distribution/ nonlinear connection / frame structure; we also
use left ''up'' and ''low'' indices as additional labels for
geometric/physical objects, for instance, in order to emphasize that $\ ^{%
\mathbf{g}}\nabla =\nabla \lbrack \mathbf{g}]$ is defined by a metric ${%
\mathbf{g}};$ the right indices are usual abstract or coordinate tensor ones.%
} For different approaches in classical and quantum gravity, there are
considered tetradic, or spinor, variables and $3+1$ spacetime decompositions
(for instance, in the so--called Arnowit--Deser--Misner, ADM, formalism,
Ashtekar variables and loop quantum gravity), or nonholonomic $2+2$
splittings, see a discussion and references in \cite{vloopdq}.

\subsection{Nonholonomic distributions and alternative connections}

For any (pseudo) Riemannian metric $\mathbf{g,}$ we can construct an
infinite number of linear connections $\ ^{\mathbf{g}}D$ which are metric
compatible, $\ ^{\mathbf{g}}D\mathbf{g}=0,$ and completely defined by
coefficients $\mathbf{g=\{\mathbf{g}_{\mu \nu }\}.}$ Of course, in general,
the torsion $\ ^{\mathbf{g}}T=\ _{D}T[\mathbf{g}]$ of a $\ ^{\mathbf{g}}D$
is not zero.\footnote{%
for a general linear connection, we do not use boldface symbols if such a
geometric object is not adapted to a prescribed nonholonomic distribution}
Nevertheless, we can work equivalently both with $^{\mathbf{g}}\nabla $ and
any $\ ^{\mathbf{g}}D,$ because the distorsion tensor $\ ^{\mathbf{g}}Z=Z[%
\mathbf{g}]$ \ from the corresponding connection deformation,
\begin{equation}
\ ^{\mathbf{g}}\nabla =\ ^{\mathbf{g}}D+\ ^{\mathbf{g}}Z,  \label{condeform}
\end{equation}%
(in the metric compatible cases, $\ ^{\mathbf{g}}Z$ is proportional to $\ ^{%
\mathbf{g}}T)$ is also completely defined by the metric structure $\mathbf{g.%
}$ In Appendix, we provide an explicit example of two metric compatible
linear connections completely defined by the same metric structure, see
formula (\ref{deflc}). Such torsions induced by nonholonomic deformations of
geometric objects are not similar to those from the Einstein--Cartan and/or
string/gauge gravity theories, where certain additional field equations (to
the Einstein equations) are considered for physical definition of torsion
fields.

Even the Einstein equations are usually formulated for the Ricci tensor and
scalar curvature defined by data $(\mathbf{g},\ ^{\mathbf{g}}\nabla ),$ the
fundamental equations and physical objects and conservation laws can be
re--written equivalently in terms of any data $(\mathbf{g,}$ $\ ^{\mathbf{g}%
}D).$ This may result in a more sophisticate structure of equations but for
well defined conditions may help, for instance, in constructing new classes
of exact solutions or to define alternative methods of quantization (like in
the Ashtekar approach to gravity) \cite{vexsol,vsgg,vrflg,vloopdq,vpla}.

In order to apply the A--model quantization via branes proposed in Ref. \cite%
{gukwit}, and relevant methods of deformation/geometric quantization, it is
convenient to select from the set of linear connections $\{\ ^{\mathbf{g}%
}D\} $ such a symplectic one which is ''canonically'' defined by the
coefficients of an Einstein metric $\mathbf{g=\{\mathbf{g}_{\mu \nu }\},}$
being compatible to a well defined almost complex and symplectic structure,
and for an associated complex manifold. In section 2 and Appendix of Ref. %
\cite{vnsak}, there are presented all details on the so--called almost K\"{a}%
hler model of general relativity (see also the constructions and
applications to Fedosov quantization of gravity in Refs. \cite%
{vpla,vegnakglfdq,vloopdq}). For convenience, we summarize in Appendix \ref%
{append1} some most important definitions and component formulas on almost K%
\"{a}hler redefinition of gravity.

Let us remember how almost K\"{a}hler variables can be introduced in
classical and quantum gravity:\quad Having prescribed on a (pseudo)
Riemannian manifold $\mathbf{V}$ a generating function $L(u)$ (this can be
any function, for certain models of analogous gravity \cite{vsgg,vrflg},
considered as a formal regular pseudo--Lagrangian $L(x,y)$ with
nondegenerate $\ ^{L}g_{ab}=\frac{1}{2}\frac{\partial ^{2}L}{\partial
y^{a}\partial y^{b}}$), we construct a canonical almost complex structure $%
\mathbf{J=}\ ^{L}\mathbf{J}$ (when $\mathbf{J\circ J=-I}$ for $\mathbf{I}$
being the unity matrix) adapted to a canonical nonlinear connection
(N--connection) structure $\mathbf{N=}\ ^{L}\mathbf{N}$ defined as a
nonholonomic distibution on $T\mathbf{V.}$ For simplicity, in this work we
shall omit left labels like $L$ if that will not result in ambiguities; it
should be emphasized that such constructions can be performed for any
regular $L,$ i.e. they do not depend explicitly on $L,$ or any local frames
or coordinates \footnote{%
We use the word ''pseudo'' because a spacetime in general relativity is
considered as a real four dimensional (pseudo) Riemannian spacetime manifold
$\mathbf{V}$ of necessary smooth class and signature $(-,+,+,+).$ For a
conventional $2+2$ splitting, the local coordinates $u=(x,y)$ on a open
region $U\subset \mathbf{V}$ are labelled in the form $u^{\alpha
}=(x^{i},y^{a}),$ where indices of type $i,j,k,...=1,2$ and $a,b,c...=3,4,$
for tensor like objects, will be considered with respect to a general
(non--coordinate) local basis $e_{\alpha }=(e_{i},e_{a}).$ One says that $%
x^{i}$ and $y^{a}$ are respectively the conventional horizontal/ holonomic
(h) and vertical / nonholonomic (v) coordinates (both types of such
coordinates can be time-- or space--like ones). \ Primed indices of type $%
i^{\prime },a^{\prime },...$ will be used for labeling coordinates with
respect to a different local basis $e_{\alpha ^{\prime }}=(e_{i^{\prime
}},e_{a^{\prime }})$ for instance, for an orthonormalized basis. For the
local tangent Minkowski space, we can chose $e_{0^{\prime }}=i\partial
/\partial u^{0^{\prime }},$ where $i$ is the imaginary unity, $i^{2}=-1,$
and write $e_{\alpha ^{\prime }}=(i\partial /\partial u^{0^{\prime
}},\partial /\partial u^{1^{\prime }},\partial /\partial u^{2^{\prime
}},\partial /\partial u^{3^{\prime }}).$}.

The canonical symplectic 1-form is defined $\mathbf{\theta (X,Y)}\doteqdot
\mathbf{g}\left( \mathbf{J}\mathbf{X,Y}\right) ,$ for any vectors $\mathbf{X}
$ and $\mathbf{Y}$ on $\mathbf{V},$ and $\ ^{\mathbf{g}}\mathbf{\theta }%
\doteqdot \ ^{L}\mathbf{\theta =}\{\mathbf{\theta }_{\mu \nu }[\mathbf{g}]\}$
and $\mathbf{g=}\ ^{L}\mathbf{g.}$ It is possible to prove by
straightforward computations that the form $\ ^{\mathbf{g}}\mathbf{\theta }$
is closed, i.e. $d\ ^{L}\mathbf{\theta }=0,$ and that there is a canonical
symplectic connection (equivalently, normal connection) $\ _{\theta }%
\widehat{\mathbf{D}}=\widehat{\mathbf{D}}=\{\ \ _{\theta }\widehat{\mathbf{%
\Gamma }}_{\ \alpha \beta }^{\gamma }=\widehat{\mathbf{\Gamma }}_{\ \alpha
\beta }^{\gamma }\}$ for which $\ \widehat{\mathbf{D}}\ ^{\mathbf{g}}\mathbf{%
\theta =}\ \widehat{\mathbf{D}}\mathbf{g=0.}$ The variables $(\ ^{\mathbf{g}}%
\mathbf{\theta },\ \widehat{\mathbf{D}})$ define an almost K\"{a}hler model
of general relativity, with distorsion of connection $\ ^{\mathbf{g}%
}Z\rightarrow \ ^{\mathbf{g}}\widehat{\mathbf{Z}}\mathbf{,}$ for $\ ^{%
\mathbf{g}}\nabla =\mathbf{\ }_{\theta }\widehat{\mathbf{D}}+\ ^{\mathbf{g}}%
\widehat{\mathbf{Z}},$ see formula (\ref{condeform}). Explicit coordinate
formulas for $\ ^{\mathbf{g}}\nabla ,\mathbf{\ }_{\theta }\widehat{\mathbf{D}%
}$ and$\ ^{\mathbf{g}}\widehat{\mathbf{Z}}$ are given by (\ref{deflc}) and (%
\ref{deft}), see proofs in Refs. \cite{vnsak,vsgg,ma2}.\ It should be noted
here that for an arbitrary nonholonomic 2+2 splitting we can construct a
Hermitian model of Einstein gravity with $d\ \mathbf{\theta }\neq 0.$ In
such a case, to perform a deformation, or other, quantization is a more
difficult problem. Choosing nonholonomic distributions generated by regular
pseudo--Lagrangians, we can work only with almost K\"{a}hler variables which
simplifies substantially the procedure of quantization, see discussions from
Refs. \cite{vpla,vfqlf,vrflg,vegnakglfdq}.

\subsection{Einstein gravity in almost K\"{a}hler variables}

In the canonical approach to the general relativity theory, one works with
the Levi--Civita connection $\bigtriangledown =\{\ _{\shortmid }\Gamma
_{\beta \gamma }^{\alpha }\}$ which is uniquely derived following the
conditions $~\ _{\shortmid }\mathcal{T}=0$ and $\bigtriangledown \mathbf{g}%
=0.$ This is a linear connection but not a distinguished connection
(d--connection) because $\bigtriangledown $ does not preserve the
nonholonomic splitting (see in Appendix the discussions relevant to (\ref%
{whitney})) under parallelism. Both linear connections $\bigtriangledown $
and $\widehat{\mathbf{D}}\equiv \ _{\theta }\widehat{\mathbf{D}}$ are
uniquely defined in metric compatible forms by the same metric structure $%
\mathbf{g}$ (\ref{gpsm}). The second one contains nontrivial d--torsion
components $\widehat{\mathbf{T}}_{\beta \gamma }^{\alpha }$ (\ref{cdtors}),
induced effectively by an equivalent Lagrange metric $\mathbf{g}=\ ^{L}%
\mathbf{g}$ (\ref{lfsm}) and adapted both to the N--connection $\ ^{L}%
\mathbf{N,}$ see (\ref{clnc}) and (\ref{whitney}), and almost symplectic $\
^{L}\mathbf{\theta }$ (\ref{asymstr}) structures $L.$

Having chosen a canonical almost symplectic d--connection, we compute the
Ricci d--tensor $\widehat{\mathbf{R}}_{\ \beta \gamma }$ (\ref{dricci}) and
the scalar curvature $\ ^{L}R$ $\ $(\ref{scalc})). Then, we can postulate in
a straightforward form the filed equations
\begin{equation}
\widehat{\mathbf{R}}_{\ \beta }^{\underline{\alpha }}-\frac{1}{2}(\
^{L}R+\lambda )\mathbf{e}_{\ \beta }^{\underline{\alpha }}=8\pi G\mathbf{T}%
_{\ \beta }^{\underline{\alpha }},  \label{deinsteq}
\end{equation}%
where $\widehat{\mathbf{R}}_{\ \ \beta }^{\underline{\alpha }}=\mathbf{e}_{\
\gamma }^{\underline{\alpha }}$ $\widehat{\mathbf{R}}_{\ \ \beta }^{\ \gamma
},$ $\mathbf{T}_{\ \beta }^{\underline{\alpha }}$ is the effective
energy--momentum tensor, $\lambda $ is the cosmological constant, $G$ is the
Newton constant in the units when the light velocity $c=1,$ and the
coefficients $\mathbf{e}_{\ \beta }^{\underline{\alpha }}$ of vierbein
decomposition $\mathbf{e}_{\ \beta }=\mathbf{e}_{\ \beta }^{\underline{%
\alpha }}\partial /\partial u^{\underline{\alpha }}$ are defined by the
N--coefficients of the N--elongated operator of partial derivation, see (\ref%
{dder}). But the equations (\ref{deinsteq}) for the canonical $\widehat{%
\mathbf{\Gamma }}_{\ \alpha \beta }^{\gamma }(\mathbf{\theta })$ are not
equivalent to the Einstein equations in general relativity written for the
Levi--Civita connection $_{\shortmid }\Gamma _{\ \alpha \beta }^{\gamma }(%
\mathbf{\theta })$ if the tensor $\mathbf{T}_{\ \beta }^{\underline{\alpha }%
} $ does not include contributions of $\ _{\shortmid }Z_{\ \alpha \beta
}^{\gamma }(\mathbf{\theta })$ in a necessary form.

Introducing the absolute antisymmetric tensor $\epsilon _{\alpha \beta
\gamma \delta }$ and the effective source 3--form
\begin{equation*}
\mathcal{T}_{\ \beta }=\mathbf{T}_{\ \beta }^{\underline{\alpha }}\ \epsilon
_{\underline{\alpha }\underline{\beta }\underline{\gamma }\underline{\delta }%
}du^{\underline{\beta }}\wedge du^{\underline{\gamma }}\wedge du^{\underline{%
\delta }}
\end{equation*}%
and expressing the curvature tensor $\widehat{\mathcal{R}}_{\ \gamma }^{\tau
}=\widehat{\mathbf{R}}_{\ \gamma \alpha \beta }^{\tau }\ \mathbf{e}^{\alpha
}\wedge \ \mathbf{e}^{\beta }$ of $\ \widehat{\mathbf{\Gamma }}_{\ \beta
\gamma }^{\alpha }=\ _{\shortmid }\Gamma _{\ \beta \gamma }^{\alpha }-\ \
_{\shortmid }\widehat{\mathbf{Z}}_{\ \beta \gamma }^{\alpha }$ as $\widehat{%
\mathcal{R}}_{\ \gamma }^{\tau }=\ _{\shortmid }\mathcal{R}_{\ \gamma
}^{\tau }-\ _{\shortmid }\widehat{\mathcal{Z}}_{\ \gamma }^{\tau },$ where $%
\ _{\shortmid }\mathcal{R}_{\ \gamma }^{\tau }$ $=\ _{\shortmid }R_{\ \gamma
\alpha \beta }^{\tau }\ \mathbf{e}^{\alpha }\wedge \ \mathbf{e}^{\beta }$ is
the curvature 2--form of the Levi--Civita connection $\nabla $ and the
distorsion of curvature 2--form $\widehat{\mathcal{Z}}_{\ \gamma }^{\tau }$
is defined by $\ \widehat{\mathbf{Z}}_{\ \beta \gamma }^{\alpha }$ (\ref%
{deft}), we derive the equations (\ref{deinsteq}) (varying the action on
components of $\mathbf{e}_{\ \beta },$ see details in Ref. \cite{vloopdq}).
The gravitational field equations are represented as 3--form equations,%
\begin{equation}
\epsilon _{\alpha \beta \gamma \tau }\left( \mathbf{e}^{\alpha }\wedge
\widehat{\mathcal{R}}^{\beta \gamma }+\lambda \mathbf{e}^{\alpha }\wedge \
\mathbf{e}^{\beta }\wedge \ \mathbf{e}^{\gamma }\right) =8\pi G\mathcal{T}%
_{\ \tau },  \label{einsteq}
\end{equation}%
\begin{eqnarray*}
\mbox{when\qquad } \mathcal{T}_{\ \tau } &=&\ ^{m}\mathcal{T}_{\ \tau }+\
^{Z}\widehat{\mathcal{T}}_{\ \tau }, \\
\ ^{m}\mathcal{T}_{\ \tau } &=&\ ^{m}\mathbf{T}_{\ \tau }^{\underline{\alpha
}}\epsilon _{\underline{\alpha }\underline{\beta }\underline{\gamma }%
\underline{\delta }}du^{\underline{\beta }}\wedge du^{\underline{\gamma }%
}\wedge du^{\underline{\delta }}, \\
\ ^{Z}\mathcal{T}_{\ \tau } &=&\left( 8\pi G\right) ^{-1}\widehat{\mathcal{Z}%
}_{\ \tau }^{\underline{\alpha }}\epsilon _{\underline{\alpha }\underline{%
\beta }\underline{\gamma }\underline{\delta }}du^{\underline{\beta }}\wedge
du^{\underline{\gamma }}\wedge du^{\underline{\delta }},
\end{eqnarray*}%
where $\ ^{m}\mathbf{T}_{\ \tau }^{\underline{\alpha }}$ is the matter
tensor field. The above mentioned equations are equivalent to the usual
Einstein equations for the Levi--Civita connection $\nabla ,$%
\begin{equation*}
\ _{\shortmid }\mathbf{R}_{\ \beta }^{\underline{\alpha }}-\frac{1}{2}(\
_{\shortmid }R+\lambda )\mathbf{e}_{\ \beta }^{\underline{\alpha }}=8\pi G\
^{m}\mathbf{T}_{\ \beta }^{\underline{\alpha }}.
\end{equation*}%
For $\widehat{\mathbf{D}}\equiv \ _{\theta }\widehat{\mathbf{D}},$ the
equations (\ref{einsteq}) define the so--called almost K\"{a}hler model of
Einstein gravity.

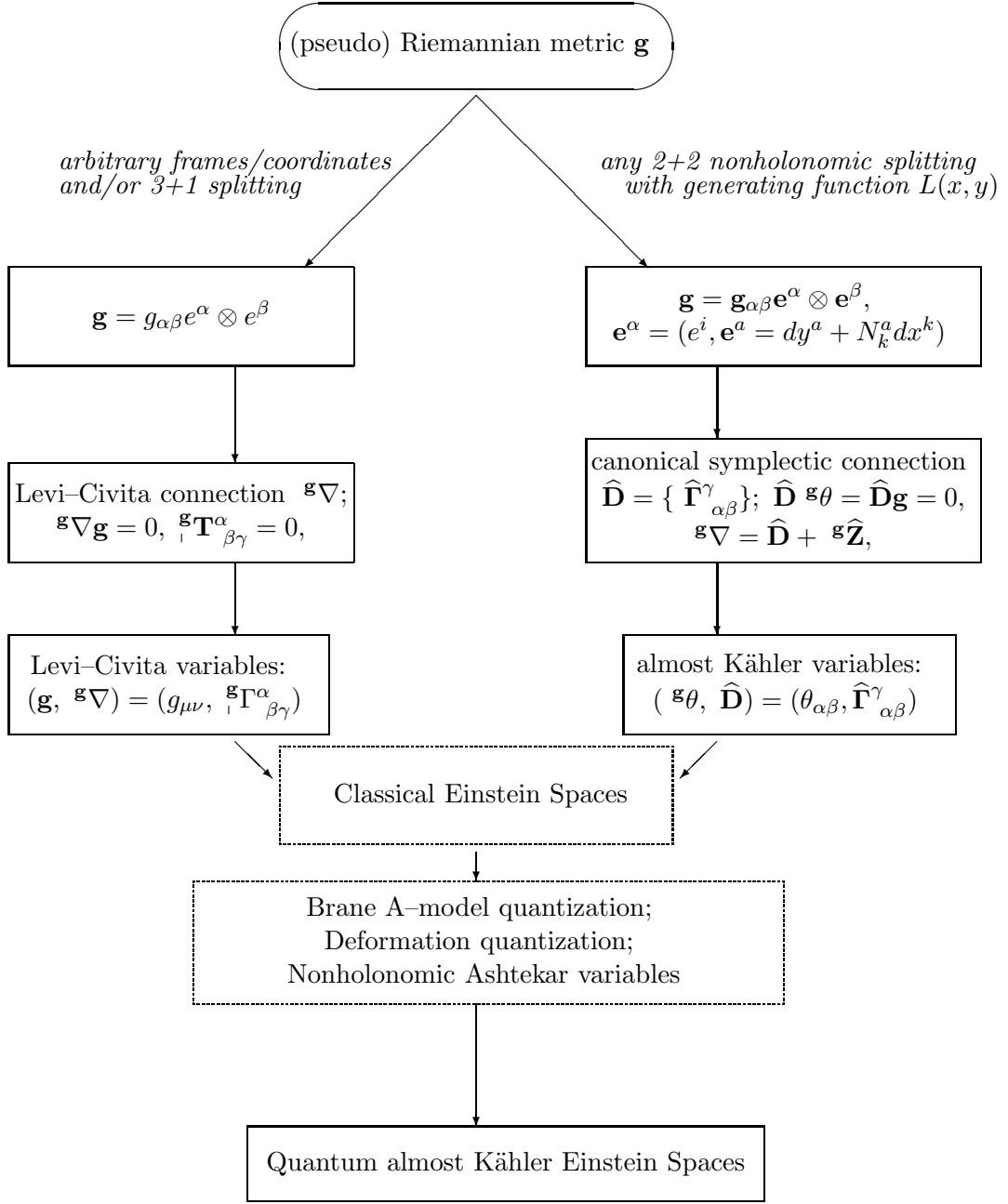
\begin{figure}[tbph]
\begin{center}
\begin{picture}(360,500)
\thinlines
\put(190,480){\oval(160,35)}
\put(114,478){\makebox{(pseudo) Riemannian metric ${\bf g}$}}
\put(20,430){\makebox{\it arbitrary frames/coordinates }}

\put(20,420){\makebox{\it and/or 3+1 splitting}}
\put(240,430){\makebox{\it any 2+2 nonholonomic splitting}}
\put(250,420){\makebox{\it with generating function $L(x,y)$}}
\put(0,350){\framebox(140,40){$\mathbf{g}=g_{\alpha \beta} e^\alpha \otimes e^\beta$}}
\put(235,350){\framebox(160,40)
{$\begin{array}{c}
  \mbox{$\mathbf{g}= \mathbf{g}_{\alpha \beta} \mathbf{e}^\alpha \otimes \mathbf{e}^\beta,$ }\\
 \mbox{$\mathbf{e}^\alpha = (e^i, \mathbf{e}^a = dy^a + N^a_k dx^k)$}
 \end{array}$
}}
\put(0,270){\framebox(140,40)
{$\begin{array}{c}
 \mbox{\ Levi--Civita connection $\ ^{\mathbf{g}}\nabla ;$}\\
 \mbox{$\ ^{\mathbf{g}}\nabla \mathbf{g}=0, \ _{\shortmid }^{\mathbf{g}}\mathbf{T}_{\ \beta \gamma }^{\alpha }=0,$}
 \end{array}$
 }}
\put(235,270){\framebox(160,50)
{$\begin{array}{c}
 \mbox{canonical symplectic connection}\\
 \mbox{$\
\widehat{\mathbf{D}}=\{\ \widehat{\mathbf{\Gamma }}_{\ \alpha \beta
}^{\gamma }\};\ \widehat{\mathbf{D}}\ ^{\mathbf{g}}\mathbf{%
\theta }= \widehat{\mathbf{D}}\mathbf{g}=0,$}\\
 \mbox{$\ ^{%
\mathbf{g}}\nabla =\widehat{\mathbf{D}}+\ ^{\mathbf{g}}%
\widehat{\mathbf{Z}},$}
 \end{array}$
 }}
 \put(0,200){\framebox(130,40)
 {$\begin{array}{c}
 \mbox{Levi--Civita variables: }\\
 \mbox{$(\mathbf{g},\ ^{\mathbf{g}}\nabla )= ( g_{\mu \nu }, \ _{\shortmid }^{\mathbf{g}}%
\Gamma _{\ \beta \gamma }^{\alpha } )$}
 \end{array}$
 }}
   \put(250,200){\framebox(135,40)
{$\begin{array}{c}
 \mbox{almost K\"{a}hler variables: }\\
 \mbox{$(\ ^{\mathbf{g}}\mathbf{\theta} ,\ \widehat{\mathbf{D}}) = (\mathbf{\theta}_{\alpha \beta}, \widehat{\mathbf{\Gamma }}_{\ \alpha \beta
}^{\gamma } ) $}
 \end{array}$
 }}
\put(110,155){\dashbox(160,40){
 \mbox{Classical Einstein Spaces}}
}
\put(75,90){\dashbox(240,50)
{$\begin{array}{c}
 \mbox{Brane A--model quantization; }\\
 \mbox{Deformation quantization; }\\
 \mbox{Nonholonomic Ashtekar variables}\\
 \end{array}$
 }}
\put(288,350){\shortstack[r]{ \vector(0,-1){30}}}
\put(92,350){\shortstack[r]{ \vector(0,-1){40}}}
\put(288,270){\shortstack[r]{ \vector(0,-1){30}}}
\put(92,270){\shortstack[r]{ \vector(0,-1){30}}}
\put(288,197){\shortstack[r]{ \vector(-1,-1){15}}}
\put(92,197){\shortstack[r]{ \vector(1,-1){15}}}
\put(190,152){\shortstack[r]{ \vector(0,-1){12}}}
\put(190,90){\shortstack[r]{ \vector(0,-1){50}}}
\put(97,10){\framebox(210,30)
{ \mbox{Quantum almost K\"{a}hler Einstein Spaces}
}}
\put(190,460){\shortstack[r]{\vector(1,-1){70}}}
\put(190,460){\shortstack[r]{\vector(-1,-1){70}}}


\end{picture}
\end{center}
\caption{\textbf{Levi--Civita and almost K\"{a}hler Variables in Gravity}}
\label{fig1}
\end{figure}

Such formulas\ expressed in terms of canonical almost symplectic form $%
\mathbf{\theta }$ (\ref{canalmsf}) and normal d--connection $\widehat{%
\mathbf{D}}\equiv \ _{\theta }\widehat{\mathbf{D}}$ (\ref{ndc}) are
necessary for encoding the vacuum field equations into cohomological
structure of quantum (in the sence of Fedosov quantization) almost K\"{a}%
hler models of the Einstein gravity, see \cite%
{vpla,vegnakglfdq,vnsak,vloopdq}.

We conclude that all geometric and classical physical information for any
data 1] $\left( \mathbf{g,}\ _{\shortmid }^{g}\Gamma \right) ,$ for Einstein
gravity, can be transformed equivalently into canonical constructions with
2] $(\ ^{\mathbf{g}}\mathbf{\theta ,\ }_{\theta }\widehat{\mathbf{D}}),$ for
an almost K\"{a}hler model of general relativity. A formal scheme for
general relativity sketching a mathematical physics ''dictionary'' between
two equivalent geometric ''languages'' (the Levi--Civita and almost K\"{a}%
hler ones) is presented in Figure \ref{fig1}.

\section{The A--Model, Quantization, and Nonholonmic Branes}

The goal of this section is to generalize the A--model approach to
quantization \cite{gukwit} for the case when branes are nonholonomic and the
symplectic structure is induced by variables $(\ ^{\mathbf{g}}\mathbf{\theta
,\ }_{\theta }\widehat{\mathbf{D}})$ in Einstein gravity.

\subsection{On quantization and nonholonomic branes}

We start with an almost K\"{a}hler model of a (pseudo) Riemannian manifold $%
\mathbf{V}$ (which is a nonholonomic manifold, or, more exactly,
N--anholonomic manifold \cite{vsgg,vrflg}, see also definitions in Appendix)
endowed with structures $(\ ^{\mathbf{g}}\mathbf{\theta ,\ }_{\theta }%
\widehat{\mathbf{D}}),$ which we wish to quantize. Our goal is to develop
the method of quantization \cite{gukwit} and to apply it to the case of
nonholonomic manifolds provided with gravitational symplectic variables. We
consider a complex line bundle $\mathcal{L}\rightarrow \mathbf{V}$ with a
unitary connection of curvature $\mathcal{R},$ like in geometric
quantization \cite{konst,sour,wood,sniat}.

In this work, we shall use an affine variety $\mathbf{Y}$ which, by
definition, is a complexification of $\mathbf{V}$ such that: 1) it is a
complex manifold with an antiholomorphic involution $\tau :\mathbf{Y}%
\rightarrow \mathbf{Y,}$ when $\mathbf{V}$ is a component of the fixed point
set of $\tau ;$ 2) there is a nondegenerate holomorphic 2--form $\ \mathbf{%
\Theta }$ on $\mathbf{Y}$ such that its restriction, $\tau ^{\ast }(\mathbf{%
\Theta }),$ to $\mathbf{V}$ is just $\ ^{\mathbf{g}}\mathbf{\theta ;}$ 3)
the unitary line bundle $\mathcal{V}\rightarrow \mathbf{V}$ can be extended
to a unitary line bundle $\mathcal{Y}\rightarrow \mathbf{Y}$ enabled with a
connection of curvature ${Re}(\mathbf{\Theta }),$ when the action of $\tau $
on $\mathbf{Y}$ results to an action on $\mathcal{Y},$ restricting to an
identity on $\mathbf{V}.$ In brief, the approach to quantization is based on
a ''good'' A--model associated with the real symplectic form $\ _{\mathbf{Y}}%
\mathbf{\theta }=Im\mathbf{\Theta }$.\footnote{%
For such a good A--model, the relevant correlation functions and observables
are complex--valued rather than formal power series depending on a formal
deformation parameter; such series converge to complex--valued functions;
following \cite{gukwit}, this is possible if the superymmetric sigma--model
with target $\mathbf{Y}$ can be twisted to give the A-model. Here, we also
note that we follow our system of denotations relevant to nonholonomic
manifolds and corresponding geometric constructions.} Such a choice
determines a canonical coisotropic brane (in our case, in general, it is a
nonholonomic manifold, because $\mathbf{V}$ is nonholonomic) in the A--model
of $\mathbf{Y}$ \cite{kapust1,kapust2}.

Similarly to \cite{gukwit}, we will make use of ordinary Lagrangian
A--branes when $\mathbf{V}$ is also modeled as a Lagrangian submanifold and
we can define a rank 1 A--brane supported on such a (nonholonomic) manifold.
We denote by $\ ^{\mathbf{g}}\mathcal{B}^{\prime }$ such a brane (one could
be inequivalent choices for a not simply--connected). It will be written $\
^{\mathbf{g}}\mathcal{B}_{cc}$ for the canonical coisotropic A--brane and $\
^{\mathbf{g}}\mathcal{B}$ for an A--brane of unspecified type. We can
construct a quantum model (i. e. quantization of $\mathbf{V}$ enabled with
variables $(\ ^{\mathbf{g}}\mathbf{\theta ,\ }_{\theta }\widehat{\mathbf{D}}%
) $) postulating that the Hilbert space associated to $\mathbf{V}$ is the
space $\ ^{\mathbf{g}}\mathcal{H}$ of $\left( \ ^{\mathbf{g}}\mathcal{B}%
_{cc},\ ^{\mathbf{g}}\mathcal{B}^{\prime }\right) .$ This is related to the
geometry of a vector bundle provided with nonlinear connection structure (in
different contexts such spaces and applications to modern physics, mechanics
and Finsler geometry and generalizations were studied in Refs. \cite%
{vsgg,vrflg,ma2}) associated to the choice of A--brane $\ ^{\mathbf{g}}%
\mathcal{B}^{\prime }.$ The first explicit constructions of such vector
spaces (without N--connection structure) were originally considered in Refs. %
\cite{aldi,gukwit}.

In order to quantize almost K\"{a}hler models of Einstein spaces, we need a
more sophisticate techniques from the so--called locally anisotropic string
theory \cite{vstr1,vstr2} and noncommutative generalizations of
Lagrange--Finsler and nononholonomic branes and gauge theories \cite%
{vncgr1,vncgr2,vncgr3,vncgr4}.\footnote{%
In low energy limits, we proved that from string/brane and/or gauge gravity
models one generates various versions of (non) commutative Lagrange--Finsler
spaces. Such constructions were considered for a long time to be very
''exotic'' and far from scopes of standard physics theories. But some years
latter, it was proven that Finsler like structures can be modelled even as
exact solutions in Einstein gravity if generic off-diagonal metrics and
nonholonomic frames are introduced into consideration \cite{vexsol,vrflg}.
More than that, the N--connection formalism formally developed in Finsler
geometry, happen to be very useful for various geometric purposes and
applications to modern gravity and quantum field theory.} This will be
discussed in further sections.

\subsection{The canonical coistoropic nonholonomic brane and A--model}

\label{ssccnb}Let us consider a complex symplectic manifold $\mathbf{Y}$
endowed with a nondegenerate holomorphic 2--form $\mathbf{\Theta }$ of type $%
(2,0)$ splitting into respective real, $\mathbf{\ }_{\mathbf{J}}\mathbf{%
\theta ,}$ and imaginary, $\ _{\mathbf{K}}\mathbf{\theta ,}$ parts, i.e.%
\begin{equation}
\mathbf{\Theta =\ }_{\mathbf{J}}\mathbf{\theta +}i\mathbf{\ }_{\mathbf{K}}%
\mathbf{\theta ,}  \label{symplhf}
\end{equation}%
where
\begin{equation}
\mathbf{I}^{t}\mathbf{\Theta }=i\mathbf{\Theta },\mbox{\ or \ }\mathbf{I}^{t}%
\mathbf{\ }_{\mathbf{J}}\mathbf{\theta =-\ }_{\mathbf{K}}\mathbf{\theta }%
\mbox{\ and \  }\mathbf{I}^{t}\mathbf{\ }_{\mathbf{K}}\mathbf{\theta }=%
\mathbf{\ }_{\mathbf{J}}\mathbf{\theta ,}  \label{auxcond2}
\end{equation}%
for $\mathbf{I}$ being the complex structure on $\mathbf{Y,}$ which may be
regarded as a linear transformation of tangent vectors, $\mathbf{I}^{t}$
denoting the transpose map acting on 1--forms; $\mathbf{\Theta }$ and $%
\mathbf{I}^{t}\mathbf{\Theta }$ are regarded as maps from tangent vectors to
1--forms.\footnote{%
even we consider such notations, we do not impose the condition that $%
\mathbf{Y}$ has a hyper--K\"{a}hler structure} In this work, we view $%
\mathbf{Y}$ as a real symplectic manifold with symplectic structure $\ _{%
\mathbf{Y}}\mathbf{\theta }=Im\mathbf{\Theta =\ }_{\mathbf{K}}\mathbf{\theta
}$ and study the associated A--model as a case in \cite{kapust1}, when such
A--models are Lagrangian branes; for our purposes, it is enough to take a $%
rank$ $1$ coisotropic A--brane whose support is just the manifold $\mathbf{Y.%
}$

Any rank 1 brane can be endowed with a unitary line bundle $\mathcal{V}$
with a connection for which we denote the curvature by $\mathbf{F}.$ If for $%
\mathbf{I}=\ _{\mathbf{Y}}\mathbf{\theta }^{-1}\ \mathbf{F,}$ we have $%
\mathbf{I}^{2}=-1,$ i.e. this is an integrable complex structure; we call
such a 1 brane to be an A--brane. We obey these conditions if we set $%
\mathbf{F}=\mathbf{\ }_{\mathbf{J}}\mathbf{\theta ,}$ when $\ _{\mathbf{Y}}%
\mathbf{\theta }^{-1}\ \mathbf{F=\ }_{\mathbf{K}}\mathbf{\theta }^{-1}\ _{%
\mathbf{J}}\mathbf{\theta }$ coincides with $\mathbf{I}$ from (\ref{auxcond2}%
). So, we can construct always an A--brane in the A--model of symplectic
structure $\mathbf{\ }_{\mathbf{Y}}\mathbf{\theta }$ starting with a complex
symplectic manifold $(\mathbf{Y,\Theta }),$ for any choice of a unitary line
bundle $\mathcal{V}$ enabled with a connection of curvature $_{\mathbf{J}}%
\mathbf{\theta} ={Re}(\mathbf{\Theta }).$ Following \cite{gukwit}, we call
this A--brane the canonical coisotropic brane and denote it as $\ ^{\mathbf{g%
}}\mathcal{B}_{cc}$ (we put a left up label $\mathbf{g}$ because for the
almost K\"{a}hler models of Einstein spaces there are induced by metric
canonical decompositions of fundamental geometric objects related to almost
complex and symplectic structures, see respective formulas (\ref{gpsm}) with
(\ref{lfsm}) and (\ref{algeq}); (\ref{acstr}) and (\ref{canalmsf})).

We emphasize that the constructions leading to an A--model depend only on $\
_{\mathbf{Y}}\mathbf{\theta }$ and do not depend on the chosen almost
complex structure; there is no need for the almost complex structure to be
integrable. So, we can always chose $_{\mathbf{Y}}\mathbf{\theta }$ to be,
let say a complexification, or proportional to the gravitational symplectic
1--form $\ ^{\mathbf{g}}\mathbf{\theta .}$ Together with the almost complex
structure $\ ^{L}\mathbf{J}$ (\ref{acstr}), this would make the A--model
more concrete (we shall consider details in section \ref{sqakeg}). Here we
adapt some key constructions from \cite{gukwit} to the case of nonholonomic
A--models, when $_{\mathbf{Y}}\mathbf{\theta =\ }_{\mathbf{K}}\mathbf{\theta
}$ $\ $for an almost complex structure $\mathbf{K}$ (in a particular case, $%
_{\mathbf{Y}}\mathbf{\theta =\ ^{\mathbf{g}}\mathbf{\theta }}$ and $\mathbf{%
K=}\ ^{L}\mathbf{J).}$ So, in this section, we consideer that $\mathbf{Y}$
is a nonholonomic complex manifold enabled with a N--connection structure
defined by a distribution of type (\ref{whitney}). For our purposes, it is
enough to consider that such a distribution is related to a decomposition of
tangent spaces of type
\begin{equation}
T\mathbf{Y}= \ T\mathbf{V}\oplus \mathbf{I}(T\mathbf{V}),  \label{whitney2}
\end{equation}%
when we chose such a $\mathbf{K}$ that $\mathbf{IK=-KI}$ implying that $%
\mathbf{J=KI}$ is also an almost complex structure.

In general, $\mathbf{J}$ and $\mathbf{K}$ are not integrable but $\mathbf{K}$
always can be defined to satisfy the properties that $\ _{\mathbf{K}}\mathbf{%
\theta }$ is of type $(1,1)$ and $\mathbf{IK}=-\mathbf{KI},$ for any
nonholonomic $\mathbf{V,}$ and the space of choices for $\mathbf{K}$ is
contractable. This can be verified for any $\mathbf{Y}$ of dimension $4k,$
for $k=1,2,...,$ as it was considered in section 2.1 \ of \cite{gukwit}. For
nonholonomic manifolds related to general relativity this is encoded not
just in properties of compact form symplectic groups, $Sp(2k),$ acting on $%
\mathbb{C}^{2k},$ and their complexification, $Sp(2k)_{\mathbb{C}}.$ The
nonholonomic $2+2$ decomposition (similarly, we can consider $n+n$), results
not in groups and Lie algebras acting on some real/complex vector spaces,
but into distinguished similar geometric objects, adapted to the
N--connection structure. In brief, they are called d--algebras, d--groups
and d--vectors. The geometry of d--groups and d--spinors, and their
applications in physics and noncommutative geometry, is considered in
details in Refs. \cite{vfspinor,vvicol,vcfalgebr}, see also Part III of \cite%
{vsgg}, and references therein (we refer readers to those works).

Here, we note that for our constructions in quantum gravity, it is enough to
take $\mathbf{Y}$ of dimension $8,$ for $k=2,$ when $\mathbf{V}$ is enabled
with a nonholonomic splitting $2+2$ and the d--group $\ ^{d}Sp(4)$ is
modelled as $\ ^{d}Sp(4)=Sp(2)\mathbf{\oplus }Sp(2)$, which is adapted to
both type decompositions (\ref{whitney}) and (\ref{whitney2}). We can now
write a sigma--model action using an associated metric $\ _{\mathbf{K}}%
\mathbf{g}=-\ _{\mathbf{K}}\mathbf{\theta}\ \mathbf{K},$ when $\mathbf{J=KI}$
will be used for quantization. The first nonhlonomic sigma-- and (super)
string models where considered in works \cite{vstr1,vstr2} for the
so--called Finsler--Lagrange (super) strings and (super) spaces, but in
those works $\mathbf{V=E,}$ for $\mathbf{E}$ being a vector (supervector
bundle). In this section, we work with complex geometric structures on
\begin{equation}
T\mathbf{Y}=h\mathbf{Y}\oplus v\mathbf{Y,}  \label{whitney3}
\end{equation}%
when such a nonholonomic splitting is induces both by (\ref{whitney2}) and (%
\ref{whitney}), i.e. $\mathbf{Y}$ is also enabled with N--connection
structure and its symplectic and complex forms are related to the
corresponding symplectic and almost complex structures on $\mathbf{V}$ (in
particular, those for the almost K\"{a}hler model of Einstein gravity).

\subsection{Space of nonholonomic $\left( \ ^{\mathbf{g}}\mathcal{B}_{cc},\
^{\mathbf{g}}\mathcal{B}^{\prime }\right) $ strings}

We can consider the space of $\left( \ ^{\mathbf{g}}\mathcal{B}_{cc},\ ^{%
\mathbf{g}}\mathcal{B}_{cc}\right) $ strings in a nonholonomic A--model as
the space of operators that can be inserted in the A--model on a boundary of
a nonholonomic string world--sheet $\mathbf{\Sigma ,}$ with a splitting $T%
\mathbf{\Sigma }=h\mathbf{\Sigma \oplus }v\mathbf{\Sigma ,}$ similarly to (%
\ref{whitney}), that ends on the brane $\ ^{\mathbf{g}}\mathcal{B}_{cc}.$ In
general, the constructions should be performed for a sigma--model with
nonholonomic target $\mathbf{Y},$ bosonic fields $\mathbf{U}$ and N--adapted
fermionic d--fields$\ _{-}\psi =\left( \ _{-}^{h}\psi ,\ _{-}^{v}\psi
\right) ,$ for left--moving d--spinors, and $_{+}\psi =\left( \ _{+}^{h}\psi
,\ _{+}^{v}\psi \right) ,$ where the $h$-- and $v$--components are defined
with respect to splitting (\ref{whitney3}).

An example of local d--operator $\mathbf{f}(\mathbf{U})$ is that
corresponding to a complex--valued function $\mathbf{f}:\mathbf{Y\rightarrow
}\mathbb{C}.$ This d--operator inserted at an interior point of $\mathbf{%
\Sigma }$ is invariant under supersymmetry transforms (on nonholonomic
supersymmetic spaces, see \cite{vstr2}) of the A--model,
\begin{equation}
\delta \mathbf{U=(}1-i\mathbf{K)}\left( \mathbf{\ }_{+}\psi \right) +\mathbf{%
(}1+i\mathbf{K)}\left( \mathbf{\ }_{-}\psi \right) ,  \label{boundoper}
\end{equation}%
if $\mathbf{f}$ is constant. The N--connection structure results in similar
transforms of the $h$- and $v$--projections of the fermionic d--fields.
Boundary d--operators must be invariant under transforms (\ref{boundoper})
of the A--model. In sigma models, one works with boundary (d-) operators,
rather than bulk (d-) operators, when the boundary conditions are
nonholonomic ones obeyed by fermionic (d-) fields,
\begin{equation*}
(\ _{\mathbf{K}}\mathbf{g-F})\left( \mathbf{\ }_{+}\psi \right) =(\ _{%
\mathbf{K}}\mathbf{g+F})\left( \mathbf{\ }_{-}\psi \right) ,
\end{equation*}%
where, for $\ _{\mathbf{K}}\mathbf{g}=-\ _{\mathbf{K}}\mathbf{\theta }\
\mathbf{K}$ and $\mathbf{F}=\mathbf{\ }_{\mathbf{J}}\mathbf{\theta ,}$ we
have $(\ _{\mathbf{K}}\mathbf{g-F})^{-1}(\ _{\mathbf{K}}\mathbf{g+F})=%
\mathbf{J=KI.}$

The boundary conditions (\ref{boundoper}) can be written in equivalent form%
\begin{eqnarray*}
\delta \mathbf{U} &\mathbf{=}&\left( \mathbf{(}1-i\mathbf{K)\ J}+\mathbf{(}%
1+i\mathbf{K)}\right) \left( \mathbf{\ }_{-}\psi \right) , \\
&=&\mathbf{(}1+i\mathbf{I)(}1+i\mathbf{J)}\left( \mathbf{\ }_{-}\psi \right)
.
\end{eqnarray*}%
This implies a topological symmetry of the nonholonomic A--model,
\begin{eqnarray}
\delta ^{1,0}\mathbf{U} &=&0\mbox{ and }\delta ^{0,1}\mathbf{U}=\mathbf{\rho
,}  \label{topolsym} \\
\delta \mathbf{\rho } &=&0,  \notag
\end{eqnarray}%
where $\mathbf{\rho }=\mathbf{(}1+i\mathbf{I)(}1+i\mathbf{J)}\left( \mathbf{%
\ }_{-}\psi \right) $ and decomposition $\delta \mathbf{U}\mathbf{=}\delta
^{1,0}\mathbf{U+}\delta ^{0,1}\mathbf{U}$ for decompositions of the two
parts of respective types (1,0) and (0,1) with respect to the complex
structure $\mathbf{I.}$ The N--splitting (\ref{whitney3}) results in $%
\mathbf{\rho }=(\ ^{h}\rho ,\ ^{v}\rho ),$ induced by h-- , v--splitting $\
_{-}\psi =\left( \ _{-}^{h}\psi ,\ _{-}^{v}\psi \right) .$

It follows from (\ref{topolsym}) that the topological supercharges of the
nonholonomic A--model corresponds to the $\overline{\partial }$ operator of $%
\mathbf{Y.}$ We wrote ''supercharges'' because one of them is for the
h--decomposition and the second one is for the v--decomposition. The
observables of the nonholonomic A--model correspond additively to the graded
d--vector space $H_{\overline{\partial }}^{0,\star }(\mathbf{Y}),$ where $%
\mathbf{Y}$ is viewed both as a complex manifold with complex structure $%
\mathbf{I}$ and as a nonholonomic manifold enabled with N--connection
structure. We shall work with the ghost number zero part of the ring of
observables which corresponds to the set of holomorphic functions on $%
\mathbf{Y.}$

For instance, all boundary observables of the nonholonomic A--model can be
constructed from $\mathbf{U}$ and $\mathbf{\rho }=(\ ^{h}\rho ,\ ^{v}\rho ).$
Let us fix the local complex coordinates on $\mathbf{Y}$ to correspond to
complex fields $\mathbf{\tilde{U}}^{\alpha }(\tau ,\sigma )=(\tilde{X}%
^{i}(\tau ,\sigma ),\tilde{Y}^{a}(\tau ,\sigma )),$ for string parameters $%
(\tau ,\sigma ),$ see also begining of section \ref{sqakeg} on coordinate
parametrizations on $\mathbf{Y}.$ \ This allows us to construct general
d--operators:

\begin{itemize}
\item of $q$--th order in $\mathbf{\rho ,}$ having d--charge $q$ under the
ghost number symmetry of the A--model, $\mathbf{\rho }^{\overline{\alpha }%
_{1}}\mathbf{\rho }^{\overline{\alpha }_{2}}...\mathbf{\rho }^{\overline{%
\alpha }_{q}}f_{\overline{\alpha }_{1}\overline{\alpha }_{2}...\overline{%
\alpha }_{q}}(\mathbf{U,}\overline{\mathbf{U}}),$ which is an d--operator as
a $(0,q)$--form on $\mathbf{Y;}$

\item of $p$--th order in $\ ^{h}\rho ,$ having h--charge $p$ under the
ghost number symmetry \ of the A--model, $\rho ^{\overline{i}_{1}}\rho ^{%
\overline{i}_{2}}...\rho ^{\overline{i}_{p}}f_{\overline{i}_{1}\overline{i}%
_{2}...\overline{i}_{p}}(\mathbf{X,}\overline{\mathbf{X}}),$ which is an
h--operator as a $(0,p)$--form on $\mathbf{Y;}$

\item of $s$-th order in $\ ^{v}\rho ,$ having v--charge $s$ under the ghost
number symmetry \ of the A--model, $\rho ^{\overline{a}_{1}}\rho ^{\overline{%
a}_{2}}...\rho ^{\overline{a}_{s}}f_{\overline{a}_{1}\overline{a}_{2}...%
\overline{a}_{s}}(\mathbf{Y,}\overline{\mathbf{Y}}),$ which is an
v--operator as a $(0,s)$--form on $\mathbf{Y.}$
\end{itemize}

For a series of consequent h- and v--operators, it is important to consider
the order of such operators.

We conclude that the constructions in this subsection determine the
d--algebra $\mathcal{A}=(\ ^{h}\mathcal{A},\ ^{v}\mathcal{A})$ of $\left( \
^{\mathbf{g}}\mathcal{B}_{cc},\ ^{\mathbf{g}}\mathcal{B}_{cc}\right) $
strings.

\subsection{Quantization for Lagrangian nonholonomic branes}

Our first purpose, in this section, is to find something (other than itself)
that the d--algebra $\mathcal{A}=(\ ^{h}\mathcal{A},\ ^{v}\mathcal{A})$ can
act on. The simplest construction is to introduce a second nonholonomic
A--brane $\ ^{\mathbf{g}}\mathcal{B}^{\prime },$ which allows us to define a
natural action of $\mathcal{A}$ on the space of $\left( \ ^{\mathbf{g}}%
\mathcal{B}_{cc},\ ^{\mathbf{g}}\mathcal{B}^{\prime }\right) $ strings. In
this paper, we consider $\ ^{\mathbf{g}}\mathcal{B}^{\prime }$ to be a
Lagrangian A--brane of rank 1, enabled with a nonholonomic distribution,
i.e. $\ ^{\mathbf{g}}\mathcal{B}^{\prime }$ is supported on a Lagrangian
nonholonomic submanifold $\mathbf{V}$ also endowed with \ a flat line bundle
$\mathcal{V}^{\prime }.$\footnote{%
The proposed interpretation of $\mathcal{V}^\prime$ is oversimplified
because of relation of branes to K--theory, possible contributions of the
the so-called B--field, disc instanton effects etc, see discussion in Ref. %
\cite{gukwit}. For the quantum gravity model to be elaborated in this paper,
this is the simplest approach. Here we also note that our system of
denotations is quite different from that by Gukov and Witten because we work
with nonholonomic spaces and distinguished geometric objects.} We assume
that $\mathbf{\ }_{\mathbf{J}}\mathbf{\theta }$ is nondegenerate when
restricted to $\mathbf{V}$ and consider $\left( \mathbf{V,\ }_{\mathbf{J}}%
\mathbf{\theta }\right) $ as a symplectic manifold to be quantized. For a
given $V,$ it is convenient to further constrain $\mathbf{K}$ such that $T%
\mathbf{V}$ is $\mathbf{J}$--invariant, when the values $\mathbf{I,K}$ and $%
\mathbf{J}=\mathbf{KI}$ obey the algebra of quaternions.

The quantization of $\left( \ ^{\mathbf{g}}\mathcal{B}_{cc},\ ^{\mathbf{g}}%
\mathcal{B}^{\prime }\right) $ strings leads to quantization of the
symplectic manifold $\left( \mathbf{V,\ }_{\mathbf{J}}\mathbf{\theta }%
\right) ,$ and in a particular case of the almost K\"{a}hler model of
Einstein gravity. We do not present a proof of this result because it is
similar to that presented in section 2.3 of \cite{gukwit} by using holonomic
manifolds and strings. For nonholonomic constructions, we have only a formal
h-- and v--component dubbing of geometric objects because of the
N--connection structure.

There are also necessary some additional constructions with the action of a
string ending on a nonholonomic brane with a Chan--Paton connection $\mathbf{%
A},$ which is given by a boundary term $\int_{\partial \mathbf{\Sigma }}%
\mathbf{A}_{\mu }d\mathbf{U}^{\mu }.$ For $\left( \ ^{\mathbf{g}}\mathcal{B}%
_{cc},\ ^{\mathbf{g}}\mathcal{B}^{\prime }\right) $ strings, we can take
that the Chan--Paton bundle $\mathcal{V}^{\prime },$ with a connection $%
\mathbf{A}^{\prime },$ of the brane $^{\mathbf{g}}\mathcal{B}^{\prime }$ is
flat but the Chan--Paton bundle of $\ ^{\mathbf{g}}\mathcal{B}_{cc}$ is the
unitary line bundle $\mathcal{V},$ with connection $\mathbf{\ A}$ of
curvature $\mathbf{\ }_{\mathbf{J}}\mathbf{\theta .}$ Next step, we consider
a line bundle $\ ^{\mathbf{g}}\mathcal{V}=\mathcal{V\otimes V}^{\prime }$
over $\mathbf{V}$ which is a unitary bundle with a connection $\mathbf{B=A-A}%
^{\prime }$ of curvature $\mathbf{\ }_{\mathbf{J}}\mathbf{\theta .}$ This
corresponds, with a corresponding approximation, to a classical action for
the zero modes%
\begin{equation*}
\int d\tau \left( \mathbf{A}_{\mu }-\mathbf{A}_{\mu }^{\prime }\right) \frac{%
du^{\mu }}{d\tau }\approx \int d\tau \mathbf{B}_{\mu }\frac{du^{\mu }}{d\tau
}.
\end{equation*}%
To quantize the zero modes with this action, and related ''quantum''
corrections, is a quantization of $\mathbf{V}$ with prequantum line bundle $%
\mathcal{V}.$

We have not yet provided a solution of the problem of quantization for the
almost K\"{a}hler model of Einstein gravity. It was only solved the second
aim to understand that if the A--model of a nonholonomic $\mathbf{Y}$
exists, then the space of $\left( \ ^{\mathbf{g}}\mathcal{B}_{cc},\ ^{%
\mathbf{g}}\mathcal{B}^{\prime }\right) $ strings can be modelled as a
result of quantizing $\mathbf{V}$ with prequantum line bundle $\ ^{\mathbf{g}%
}\mathcal{V}.$ It is still difficult to describe such spaces explicitly, in
a general case, but the constructions became well defined for $(\ ^{\mathbf{g%
}}\mathbf{\theta ,\ }_{\theta }\widehat{\mathbf{D}})$ on $\mathbf{V.}$ Using
such information from the geometry of almost K\"{a}hler spaces, induced by a
generating function with nonholonomic 2+2 splitting, we can establish
certain important general properties of the A--model to learn general
properties of quantization.

\subsubsection{Properties of distinguished Hilbert spaces for nonholonomic
A--models}

There are two interpretations of the space of $\left( \ ^{\mathbf{g}}%
\mathcal{B}_{cc},\ ^{\mathbf{g}}\mathcal{B}^{\prime }\right) $ strings: The
first one is to say that there is a Hilbert space $\ ^{\mathbf{g}}\mathcal{H}
$ in the space of such strings in the nonholonomic A--model of $\ _{\mathbf{Y%
}}\mathbf{\theta} = \ _{\mathbf{K}}\mathbf{\theta .}$ The second one is to
consider a Hilbert space $\ ^{\mathbf{g}}\widetilde{\mathcal{H}}$ of such
strings in the B--model of complex structure $\mathbf{J}.$ For a compact $%
\mathbf{V,}$ or for wave functions required to vanish sufficiently rapidly
at infinity, both spaces $\mathcal{H}$ and $\widetilde{\mathcal{H}}$ are the
same: they can be described as the space of zero energy states of the
sigma--model with target $\mathbf{Y.}$ The model is compactified on an
interval with boundary conditions at the ends determined by nonholonomic $\
^{\mathbf{g}}\mathcal{B}_{cc}$ and$\ ^{\mathbf{g}}\mathcal{B}^{\prime }.$ It
should be noted here that both $\ ^{\mathbf{g}}\mathcal{H}$ and $\ ^{\mathbf{%
g}}\widetilde{\mathcal{H}}$ are $Z$--graded by the ''ghost number'' but
differently, such grading being conjugate, because the A---model is of type $%
\mathbf{\ }_{\mathbf{K}}\mathbf{\theta }$ and the B--model is of type $%
\mathbf{\ }_{\mathbf{Y}}\mathbf{\theta .}$

We can use a sigma--model of target $\mathbf{Y}$ when the boundary
conditions are set by two branes of type $(\mathbf{A,B,A}).$ In this case,
the model has an $SU(2)$ group of so--called R--symmetries, called $%
SU(2)_{R},$ with two ghost number symmetries being conjugate but different $%
U(1)$ subgroups of $SU(2)_{R}.$ For nonholonomic models, we have to dub the
groups and subgroups correspondingly for the h-- and v--subspaces. In the
simplest case, we can perform quantization for trivial grading, i.e. when $\
^{\mathbf{g}}\mathcal{V}$ is very ``ample`` as a line bundle in complex
structure $\mathbf{J}.$

For the B--model, we can chose any K\"{a}hler metric, for instance, to
rescale the metric of $\mathbf{Y}$ as to have a valid sigma--model
perturbation theory. In such a limit, it is possible to describe $\ ^{%
\mathbf{g}}\widetilde{\mathcal{H}}$ by a $\overline{\partial }$ cohomology,%
\begin{equation*}
\ ^{\mathbf{g}}\widetilde{\mathcal{H}}=\oplus _{\underline{j}=0}^{\dim _{%
\mathbb{C}}\mathbf{V}}H^{\underline{j}}(\mathbf{V,}\mathcal{K}^{1/2}\otimes
\ ^{\mathbf{g}}\mathcal{V}),
\end{equation*}%
with a similar decomposition for $\ ^{\mathbf{g}}\mathcal{H}\cong \ ^{%
\mathbf{g}}\widetilde{\mathcal{H}},$ i.e.
\begin{equation}
\ ^{\mathbf{g}}\mathcal{H}=\oplus _{\underline{j}=0}^{\dim _{\mathbb{C}}%
\mathbf{V}}H^{\underline{j}}(\mathbf{V,}\mathcal{K}^{1/2}\otimes \ ^{\mathbf{%
g}}\mathcal{V}).  \label{hermspace}
\end{equation}%
The value $\mathcal{K}^{1/2}$ is the square root (this is a rough
approximation) of the canonical line bundle $\mathcal{K}$ on $\mathbf{V.}$
This is because, in general, $\mathbf{V}$ may not be a spin manifold. Such
nonholonomic configurations related to gerbes are discussed in \cite%
{vgerb1,vgerb2}. In relation to $K$--theory, details are given in \cite%
{gukwit} (nonholonomic configurations existing in our constructions do not
change those conclusions). Here, we note that for very ample $\ ^{\mathbf{g}}%
\mathcal{V}$ \ the cohomology vanishes except for $\underline{j}=0$ and its $%
\mathbb{Z}$--grading is trivial. This is one problem. The second limitation
is that $\ ^{\mathbf{g}}\mathcal{H}$ is described as a vector space which
does not lead to a natural description as a Hilbert space with a hermitian
inner product. This description of $\ ^{\mathbf{g}}\mathcal{H}$ has certain
resemblance to constructions in geometric quantization because the above
cohomology defines quantization with a complex polarization. Nevertheless,
this paper is based on the Gukov--Witten approach to quantization and does
not provide a variant of geometric quantization, see in \cite{gukwit} why
this is not geometric quantization.

\subsubsection{Topological restrictions and unitarity}

There is a topological obstruction to having a $Spin_{c}$ structure and
there are further obstructions to having a flat $Spin_{c}$ structure on $%
\mathbf{V}.$ Such nonholonomic spinor constructions were analyzed firstly in
relation to definition of Finsler--Lagrange spinors \cite{vfspinor}, further
developments are outlined in Refs. \cite{vvicol,vcfalgebr}. The problem of
definition of spinors and Dirac operators for nonholonomic manifolds can be
solved by the same N--connection methods with that difference that instead
of vector/tangent bundles we have to work with manifolds enabled with
nonholonomic distributions.

In general, $Spin_{c}$ structures on $\mathbf{V}$ are classified
topologically: we have to chose a way of lifting the second
Stieffel--Whitney class $w_{2}(\mathbf{V})$ $\in H^{2}(\mathbf{V,}\mathbb{Z}%
_{2})$ to an integral cohomology class $\zeta \in H^{2}(\mathbf{V,}\mathbb{Z}%
_{2}).$ In their turn, flat $Spin_{c}$ structures are parametrized and
classified by a choice of a lift $\zeta $ as a torsion element of $H^{2}(%
\mathbf{V,}\mathbb{Z}_{2}).$ We emphasize that a sympletic manifold that
does not admit a flat $Spin_{c}$ structure cannot be quantized in the sense %
\cite{gukwit}. Perhaps, the cohomological analysis used in Fedosov
quantization for almost K\"{a}hler models of gravity \cite%
{vpla,vfqlf,vegnakglfdq}, can be re--defined for the Gukov--Witten approach.
In this work, we shall consider such gravitational fields and their
quantization when the flat $Spin_{c}$ structure exists and they are
distinguished into h-- and v--components adapted to N--connection structures
defined by certain generating functions.

Our next purpose is to define a Hermitian inner product on $\ ^{\mathbf{g}}%
\mathcal{H}$ (\ref{hermspace}). For our further application in quantum
gravity, we chose a nonholonomic A--model as a twisted version of a standard
physical model, unitary, defined also as a supersymmetric field theory. Such
a theory has an antilinear CPT symmetry, in our approach denoted $\mathbf{%
\Xi }.$ This operator maps any $\left( \mathcal{B}_{1},\ \mathcal{B}%
_{2}\right) $ string into a $\left( \mathcal{B}_{2},\ \mathcal{B}_{1}\right)
$ which also defines an antilinear map from $\left( \ ^{\mathbf{g}}\mathcal{B%
}_{cc},\ ^{\mathbf{g}}\mathcal{B}^{\prime }\right) $ strings into $\left( \
^{\mathbf{g}}\mathcal{B}^{\prime },\ ^{\mathbf{g}}\mathcal{B}_{cc}\right) $
strings, but this is not a symmetry of an A--model. In explicit form, the
definition of an A--model depends on a choice of a differential $\mathbf{Q}$
as a complex linear combination of supercharges. For nonholonomic
configurations, we work with couples of h-- and v--supercharges $\mathbf{Q=}%
(\ ^{h}Q,\ \ ^{v}Q).$ The maps with CPT symmetry transform $\mathbf{Q}$ into
its Hermitian adjoint $\mathbf{Q}^{+}\mathbf{=}(\ ^{h}Q^{+},\ \ ^{v}Q^{+})$
being the differential of a complex conjugate nonholonomic A--model.

Let us suppose that the nonholonomic complex manifold $\mathbf{Y}$ admits an
involution $\tau ,$ i.e. a N--adapted diffeomorphism obeying $\tau ^{2}=1,$
with the (odd) property:%
\begin{equation}
\tau ^{\ast }(\mathbf{\ }_{\mathbf{K}}\mathbf{\theta )=-\ }_{\mathbf{K}}%
\mathbf{\theta .}  \label{oddpropr}
\end{equation}%
Such an operator can be always introduced on $\mathbf{Y}$ by construction.
This $\tau $ maps a nonholonomic A--model into a conjugate nonholonomic
A--model and $\mathbf{\Xi }_{\tau }=\tau \mathbf{\Xi }$ is a N--adapted
antilinear map from the nonholonomic A--model to itself. This is a general
property which holds both for holonomic and nonholonomic geometrical models
of certain underlying physical theory; one may be a twisting of the
underlying model and in such a case the structure $\mathbf{K}$\textbf{\ }can
be chosen to be integrable.

We consider explicitly the antiholomorphic involution $\tau $ for a
nonholonomic complex symplectic manifold defined by data $(\mathbf{%
Y,I,\Theta }),$ with a lift to $\mathcal{V},$ where $\mathbf{\Theta =\ }_{%
\mathbf{J}}\mathbf{\theta +}i\mathbf{\ }_{\mathbf{K}}\mathbf{\theta ,I=\ }_{%
\mathbf{J}}\mathbf{\theta }^{-1}\mathbf{\ }_{\mathbf{K}}\mathbf{\theta ,}$
for $\mathbf{\ }_{\mathbf{J}}\mathbf{\theta }$ being the curvature of the
Chan--Paton bundle of the ($\tau $--invariant) brane $\ ^{\mathbf{g}}%
\mathcal{B}_{cc},$ when
\begin{equation*}
\tau ^{\ast }(\mathbf{\ }_{\mathbf{J}}\mathbf{\theta )=\ }_{\mathbf{J}}%
\mathbf{\theta ,\ }\tau ^{\ast }(\mathbf{I)}=-\mathbf{I},\mathbf{\ }\tau
^{\ast }(\mathbf{\Theta )}=\overline{\mathbf{\Theta }}\mathbf{.}
\end{equation*}%
Using the topological inner product $(\ ,\ )$ as the pairing between (in
general, nonholonomic) $\left( \mathcal{B}_{1},\ \mathcal{B}_{2}\right) $
strings and $\left( \mathcal{B}_{2},\ \mathcal{B}_{1}\right) ,$ we can
introduce the inner product on $^{\mathbf{g}}\mathcal{H}$ as
\begin{equation*}
<\psi ,\psi ^{\prime }>=<\mathbf{\Xi }_{\tau }\psi ,\psi ^{\prime }>.
\end{equation*}%
So, we conclude that if $\mathcal{B}_{1}\ $and $\mathcal{B}_{2}$ are $\tau $%
--invariant nonholonomic A--branes, we can use $\mathbf{\Xi }_{\tau }$ to
define a Hermitian inner product on the (already Hilbert) \ space $^{\mathbf{%
g}}\mathcal{H}$ of $\left( \mathcal{B}_{1},\ \mathcal{B}_{2}\right) $
strings.

\subsubsection{A Hermitian inner product on $\ ^{\mathbf{g}}\mathcal{H}$ of $%
\left( \ ^{\mathbf{g}}\mathcal{B}_{cc},\ ^{\mathbf{g}}\mathcal{B}%
^{\prime}\right) $ strings}

We suppose that the nonholonomic spacetime manifold $\mathbf{V}$ is $\tau $%
--invariant and that there is a lift of $\tau $ to act on the Chan--Paton
line bundle $\mathcal{V}^{\prime }$ on $\mathbf{V.}$ In such cases, the
corresponding nonholonomic Lagrangian A--brane $\ ^{\mathbf{g}}\mathcal{B}%
^{\prime }$ is also $\tau $--invariant. This allows us to construct a
Hermitian form, and corresponding inner product $<\ ,\ >,$ on the space $^{%
\mathbf{g}}\mathcal{H}$ of $\left( \ ^{\mathbf{g}}\mathcal{B}_{cc},\ ^{%
\mathbf{g}}\mathcal{B}^{\prime }\right) $ strings if $\tau $ maps $\mathbf{V}
$ to itself.

Nearly the classical limit, the norm of a state \ $\psi \in \ ^{\mathbf{g}}%
\mathcal{H}$ \ is approximated \ $<\psi ,\psi >=\int_{\mathbf{V}}\overline{%
\psi }(\tau u)\psi (u)\delta u;$ this form is positive--definite only if $%
\tau $ acts trivially on $\mathbf{V.}$ In general, $<\psi ,\psi >$ is
nondegenerate but not necessarily positive definite. These are some
consequences from nondegeneracy of topological inner product $(\ ,\ )$ and
the property that $\mathbf{\Xi }_{\tau }^{2}=1.$ We can chose any $\psi _{0}$
such that $(\psi _{0},\psi )\neq 0$ and set $\psi _{1}=\mathbf{\Xi }_{\tau
}\psi _{0}$ which results in $<\psi _{1},\psi >\neq 0,$ i.e. nondegenerate
property, but this dos not constrain that $<\psi ,\psi >\ >0$ for all
nonzero $\psi \in \ ^{\mathbf{g}}\mathcal{H}.$ For the inverse construction,
we consider $\mathbf{V}$ to be a component of the fixed point set of map $%
\tau .$ Because of property (\ref{oddpropr}), we get that $\mathbf{V}$ is
Lagrangian for $\mathbf{\ }_{\mathbf{K}}\mathbf{\theta }$ and that $\mathbf{%
\ }_{\mathbf{J}}\mathbf{\theta }$ is nondegenerate when restricted to $%
\mathbf{V}.$ Such properties are automatically satisfied for the almost K%
\"{a}hler model of Einstein gravity with splitting (\ref{whitney3}). $\ $The
map $\tau $ acts as 1 and -1 $\ $\ on summands $T\mathbf{V}$ and $\mathbf{I}T%
\mathbf{V}$ and $\mathbf{\ }_{\mathbf{J}}\mathbf{\theta }$ is the sum of
nondegenerate 2--forms on respective spaces.

There is a construction leading to the classical limit \cite{gukwit}, even
in our case we have certain additional nonholonomic distributions. For this,
we can take the space of $\left( \ ^{\mathbf{g}}\mathcal{B}_{cc},\ \ ^{%
\mathbf{g}}\mathcal{B}_{cc}\right) $ strings and perform the deformation
quantization of the complex nonholonomic symplectic manifold $\mathbf{Y}$
(such constructions are presented in detail for almost complex models of
gravity \cite{vpla,vfqlf,vegnakglfdq}). We get an associative d--algebra $\
^{\mathbf{g}}\mathcal{A}.$ Then we chose an antiholomorphic involution $\tau
,$ with a lift to the line bundle $\mathcal{Y}\rightarrow \mathbf{Y,}$ and a
component $\mathbf{V}$ of the fixed point set supporting a $\tau $%
--invariant nonholonomic A--brane $\ ^{\mathbf{g}}\mathcal{B}^{\prime }.$
Now, we can say that $\ ^{\mathbf{g}}\mathcal{A}$ acts on $^{\mathbf{g}}%
\mathcal{H}$ defined as the space of $\left( \ ^{\mathbf{g}}\mathcal{B}%
_{cc},\ \ ^{\mathbf{g}}\mathcal{B}^{\prime }\right) $ strings. To consider
the action $\mathbf{\Xi }_{\tau }$ we model such a map acts on a function on
$\mathbf{Y,}$ defining a $\left( \ ^{\mathbf{g}}\mathcal{B}_{cc},\ \ ^{%
\mathbf{g}}\mathcal{B}_{cc}\right) $ string as the composition of $\tau $
with complex conjugation. For an operator $\mathcal{O}_{f}:\ ^{\mathbf{g}}%
\mathcal{H\rightarrow }\ ^{\mathbf{g}}\mathcal{H}$ associated to a function $%
f,$ we define the Hermitian adjoint of $\mathcal{O}_{f}$ to be associated
with the function $\tau (\overline{f}).$ Working with a real function $f$ \
when restricted to $\mathbf{V}$ and if $\tau $ leaves $\mathbf{V}$ fixed
pointwise, we get $\tau (\overline{f})=f$ with a Hermitian $\mathcal{O}_{f}.$

Finally, in this section, we conclude that the Gukov--Witten method \cite%
{gukwit} really allows us to construct a physical viable Hilbert space for
quantum almost K\"{a}hler models of Einstein gravity related to the Fedosov
quantization of a corresponding complex nonholonomic symplectic manifold $%
\mathbf{Y.}$

\section{Quantization of the Almost K\"{a}hler Model of Einstein Gravity}

\label{sqakeg}In this section, we provide explicit constructions for quantum
physical states of the almost K\"{a}hler model of Einstein gravity.

\subsection{Coordinate parametrizations for almost K\"{a}hler gravitational
A--models}

\label{sskgm}The local coordinates on a nonholonomic complex manifold $%
\mathbf{Y}$ are denoted $u^{\widetilde{\alpha }}=(u^{\alpha },iu^{\grave{%
\alpha}}),$ for $i^{2}=-1,$ where $u^{\alpha }=(x^{i},y^{a})$ are local
coordinates on $\mathbf{V},$ and $\grave{u}^{\alpha }=u^{\grave{\alpha}}=(%
\grave{x}^{i}=x^{\grave{\imath}},\grave{y}^{a}=y^{\grave{a}})$ are real
local (pseudo) Euclidean coordinates of a conventional ''left--primed''
nonholonomic manifold $\mathbf{\grave{V}.}$ We emphasize that ''nonprimed''
indices can not be contracted with primed indices, because they label
objects on different spaces. For (non)holonomic Einstein spaces, the
coordinate indices will run values $i,j,...=1,2;a,b,...=3,4$ and $\grave{%
\imath},\grave{j},...=1,2;\grave{a},\grave{b},...=3,4.$ We can also treat $%
\tilde{u}^{\alpha }=$ $u^{\widetilde{\alpha }}=(\tilde{x}^{j}=x^{j}+i\grave{x%
}^{j},\tilde{y}^{a}=y^{a}+i\grave{y}^{a})$ as complex coordinates on $%
\mathbf{Y,}$ when, for instance, $\tilde{u}^{1}=u^{1}+i\grave{u}^{1}.$ In
brief, such coordinates are labelled respectively $\tilde{u}=(\tilde{x},%
\tilde{y}),u=(x,y)$ and $\grave{u}=(\grave{x},\grave{y}).$

In general, such real manifolds $\mathbf{V,\grave{V}}$ and complex manifold $%
\mathbf{Y}$ are enabled with respective N--connection structures, see
formula (\ref{coeffnc}), $\mathbf{N}=\{N_{i}^{a}(u)\}\mathbf{,\grave{N}}=\{%
\grave{N}_{i}^{a}=N_{\grave{\imath}}^{\grave{a}}(\grave{u})\}$ and $\mathbf{%
\tilde{N}}=\{\tilde{N}_{i}^{a}=N_{\tilde{\imath}}^{\tilde{a}}(\tilde{u}%
)\}=\{N_{i}^{a}(\tilde{u})+i\grave{N}_{i}^{a}(\tilde{u})\}.$ For $\mathbf{%
\grave{V}}$ and $\mathbf{Y,}$ the corresponding N--adapted bases (\ref{dder}%
) are%
\begin{eqnarray*}
\mathbf{\grave{e}}_{\nu } &=&\left( \mathbf{\grave{e}}_{i}=\frac{\partial }{%
\partial \grave{x}^{i}}-\grave{N}_{i}^{a}(\grave{u})\frac{\partial }{%
\partial \grave{y}^{a}},\grave{e}_{a}=\frac{\partial }{\partial \grave{y}^{a}%
}\right) = \\
\mathbf{e}_{\grave{\nu}} &=&\left( \mathbf{e}_{\grave{\imath}}=\frac{%
\partial }{\partial x^{\grave{\imath}}}-N_{\grave{\imath}}^{\grave{a}}(%
\grave{u})\frac{\partial }{\partial y^{\grave{a}}},e_{\grave{a}}=\frac{%
\partial }{\partial y^{\grave{a}}}\right)
\end{eqnarray*}%
and
\begin{equation*}
\mathbf{\tilde{e}}_{\nu }=\left( \mathbf{\tilde{e}}_{i}=\frac{\partial }{%
\partial \tilde{x}^{i}}-\tilde{N}_{i}^{a}(\tilde{u})\frac{\partial }{%
\partial \tilde{y}^{a}},\tilde{e}_{a}=\frac{\partial }{\partial \tilde{y}^{a}%
}\right)
\end{equation*}%
and dual bases (\ref{ddif}) are
\begin{eqnarray*}
\mathbf{\grave{e}}^{\mu } &=&\left( \grave{e}^{i}=d\grave{x}^{i},\mathbf{%
\grave{e}}^{a}=d\grave{y}^{a}+\grave{N}_{i}^{a}(\grave{u})d\grave{x}%
^{i}\right) = \\
\mathbf{e}^{\grave{\mu}} &=&\left( e^{\grave{\imath}}=dx^{\grave{\imath}},%
\mathbf{e}^{\grave{a}}=dy^{\grave{a}}+N_{\grave{\imath}}^{\grave{a}}(%
\grave{u})dx^{\grave{\imath}}\right)
\end{eqnarray*}%
and
\begin{equation*}
\mathbf{\tilde{e}}^{\mu }=\left( \tilde{e}^{i}=d\tilde{x}^{i},\mathbf{\tilde{%
e}}^{a}=d\tilde{y}^{a}+\tilde{N}_{i}^{a}(\tilde{u})d\tilde{x}^{i}\right) .
\end{equation*}%
We can also use parametrizations
\begin{equation*}
\mathbf{\tilde{e}}_{\nu }=\mathbf{e}_{\tilde{\nu}}=\left(
\begin{array}{c}
\mathbf{e}_{\nu } \\
i\mathbf{e}_{\grave{\nu}}%
\end{array}%
\right) \in T\mathbf{Y\mbox{\ and \ }\tilde{e}}^{\mu }=\mathbf{e}^{\tilde{\mu%
}}=\left( \mathbf{e}^{\mu },-i\mathbf{\grave{e}}^{\mu }\right) \in T^{\ast }%
\mathbf{Y.}
\end{equation*}%
The above presented formulas for N--adapted (co) bases are derived following
splitting (\ref{whitney2}) and (\ref{whitney3}).

The almost complex structure $\mathbf{K}$ on $\mathbf{\grave{V}}$ is defined
similarly to (\ref{acstr})
\begin{equation*}
\mathbf{K}=\mathbf{K}_{\ \beta }^{\alpha }\ \grave{e}_{\alpha }\otimes
\grave{e}^{\beta }=\mathbf{K}_{\ \underline{\beta }}^{\underline{\alpha }}\
\frac{\partial }{\partial \grave{u}^{\underline{\alpha }}}\otimes d\grave{u}%
^{\underline{\beta }}=\mathbf{K}_{\ \beta ^{\prime }}^{\alpha ^{\prime }}\
\mathbf{\grave{e}}_{\alpha ^{\prime }}\otimes \mathbf{\grave{e}}^{\beta
^{\prime }}=\mathbf{-}\grave{e}_{2+i}\otimes \grave{e}^{i}+\mathbf{\grave{e}}%
_{i}\otimes \ \mathbf{\grave{e}}^{2+i}.
\end{equation*}

The almost symplectic structure on a manifold $\mathbf{V}$ is defined by $%
\mathbf{\ }_{\mathbf{J}}\mathbf{\theta =\ ^{\mathbf{g}}\mathbf{\theta }}$ $%
\mathbf{\mathbf{=}\theta }=\ ^{L}\mathbf{\theta ,}$ see formula (\ref%
{canalmsf}). A similar construction can be defined on $\mathbf{\grave{V}}$
as $\mathbf{\ }_{\mathbf{K}}\mathbf{\theta =\ ^{\mathbf{\grave{g}}}\mathbf{%
\grave{\theta}=}\grave{\theta}}=\ ^{\grave{L}}\mathbf{\grave{\theta}}$
\begin{eqnarray*}
\mathbf{\grave{\theta}} &=&\ ^{\grave{L}}\mathbf{\grave{\theta}}=\frac{1}{2}%
\ ^{\grave{L}}\mathbf{\grave{\theta}}_{ij}(\grave{u})\grave{e}^{i}\wedge
\grave{e}^{j}+\frac{1}{2}\ \ ^{\grave{L}}\mathbf{\grave{\theta}}_{ab}(%
\grave{u})\mathbf{\grave{e}}^{a}\wedge \mathbf{\grave{e}}^{b} \\
&=&\grave{g}_{ij}(\grave{x},\grave{y})\left[ d\grave{y}^{i}+\grave{N}%
_{k}^{i}(\grave{x},\grave{y})d\grave{x}^{k}\right] \wedge d\grave{x}^{j}.
\end{eqnarray*}%
Here, it should be noted that, in general, we can consider different
geometric objects like the generation function $L(x,y),$ and metric $%
g_{ij}(x,y)$ on $\mathbf{V}$ and, respectively, $\grave{L}(\grave{x},\grave{y%
})$ and $\grave{g}_{ij}(\grave{x},\grave{y})$ on $\mathbf{\grave{V}},$ (as
some particular cases, we can take two different exact solutions of
classical Einstein equations). But the constructions from this section hold
true also if we dub identically the geometric constructions on $\mathbf{V}$
and $\mathbf{\grave{V}.}$

A canonical holomorphic 2--form $\mathbf{\Theta }$ of type $(2,0)$ $\ $\ on $%
\mathbf{Y,}$ see (\ref{symplhf}), induced from the almost K\"{a}hler model
of Einstein gravity, is computed
\begin{equation*}
\mathbf{\Theta =\Theta }_{\tilde{\mu}\tilde{\nu}}\mathbf{\ e}^{\tilde{\mu}%
}\wedge \mathbf{e}^{\tilde{\nu}},
\end{equation*}%
for $\mathbf{\Theta }_{\tilde{\mu}\tilde{\nu}}=\mathbf{\tilde{\Theta}}_{\mu
\nu }=\left(
\begin{array}{cc}
_{\mathbf{J}}\mathbf{\theta }_{\mu \nu }(\tilde{u}) & 0 \\
0 & -i\mathbf{\ }_{\mathbf{K}}\mathbf{\theta }_{\mu \nu }(\tilde{u})%
\end{array}%
\right) $ and general complex coordinates $\tilde{u}^{\mu }.$ This form is
holomorphic by construction, $\overline{\partial }\mathbf{\Theta =0,}$ for $%
\overline{\partial }$ defined by complex adjoints of $\tilde{u}^{\mu }.$

The above presented formulations allow us a straightforward redefinition of
the component tensor calculus adapted to N--connections structures on real
nonholonomic manifolds and their almost K\" ahler models (see Appendix and
details in Refs. \cite{vsgg,vrflg,vegnakglfdq,vncgr4,vcfl2} ) to similar
constructions with complex nonholonomic manifolds and geometric objects on
such complex spaces.

\subsection{N--adapted symmetries for nonholonomic curve flows and
bi--Hamilton structures}

An explicit construction of a Hilber space $\ ^{\mathbf{g}}\mathcal{H}$ with
a Hermitian inner product on $\left( \ ^{\mathbf{g}}\mathcal{B}_{cc},\ ^{%
\mathbf{g}}\mathcal{B}^{\prime }\right) $ strings for the almost K\"{a}hler
model of Einstein gravity is possible if we prescribe in the theory certain
generic groups of symmetries. There are proofs that metric structures on a
(pseudo) Riemannian manifold \cite{vcfl1,vcfl2} can be decomposed into
solitonic data with corresponding hierarchies of nonlinear waves. Such
constructions hold true for more general types of
Finsler--Lagrange--Hamilton geometries and/or their Ricci flows \cite%
{vcflrf3,vcfanco4} and related to the geometry of curve flows adapted to a
N--connection structure on a (pseudo) Riemannian (in general, nonholonomic)
manifold $\mathbf{V.}$\footnote{%
A non--stretching curve $\gamma (\tau ,\mathbf{l})$ on $\mathbf{V,}$ where $%
\tau $ is a real parameter and $\mathbf{l}$ is the arclength of the curve on
$\mathbf{V,}$ is defined with such evolution d--vector $\mathbf{Y}=\gamma
_{\tau }$ and tangent d--vector $\mathbf{X}=\gamma _{\mathbf{l}}$ that $%
\mathbf{g(X,X)=}1\mathbf{.}$ Such a curve $\gamma (\tau ,\mathbf{l})$ swept
out a two--dimensional surface in $T_{\gamma (\tau ,\mathbf{l})}\mathbf{V}%
\subset T\mathbf{V.}$} Our idea is to consider in quantum gravity models the
same symmetries as for the ''solitonic'' encoding of classical gravitational
interactions.

A well known class of Riemannian manifolds for which the frame curvature
matrix constant consists of the symmetric spaces $M=G/H$ for compact
semisimple Lie groups $G\supset H.$ A complete classification and summary of
main results on such spaces are given in Refs. \cite{helga,kob}. The class
of nonholonomic manifolds enabled with N--connection structure are
characterized by conventional nonholonomic splitting of dimensions. For
explicit constructions, we suppose that the ''horizontal'' distribution is a
symmetric space $hV=hG/SO(n)$ with the isotropy subgroup $hH=SO(n)\supset
O(n)$ and the typical fiber space is a symmetric space $F=vG/SO(m)$ with the
isotropy subgroup $vH=SO(m)\supset O(m).$ This means that $hG=SO(n+1)$ and $%
vG=SO(m+1)$ which is enough for a study of real holonomic and nonholonomic
manifolds and geometric mechanics models.\footnote{%
we can consider $hG=SU(n)$ and $vG=SU(m)$ for geometric models with spinor
and gauge fields and in quantum gravity}

The Riemannian curvature and the metric tensors for $M=G/H$ are covariantly
constant and $G$--invariant resulting in constant curvature matrices. Such
constructions are related to the formalism of bi--Hamiltonian operators,
originally investigated for symmetric spaces with $M=G/SO(n)$ with $%
H=SO(n)\supset O(n-1)$ and when $G=SO(n+1),SU(n),$ see \cite{sharpe} and
references in \cite{vcflrf3,vcfanco4}.

For nonholonomic manifolds, our aim was to solder in a canonic way (like in
the N--connection geometry) the horizontal and vertical symmetric Riemannian
spaces of dimension $n$ and $m$ with a (total) symmetric Riemannian space $V$
of dimension $n+m,$ when $V=G/SO(n+m)$ with the isotropy group $%
H=SO(n+m)\supset O(n+m)$ and $G=SO(n+m+1).$ There are natural settings to
Klein geometry of the just mentioned horizontal, vertical and total
symmetric Riemannian spaces: The metric tensor $hg=\{\mathring{g}_{ij}\}$ on
$h\mathbf{V}$ is defined by the Cartan--Killing inner product $<\cdot ,\cdot
>_{h}$ on $T_{x}hG\simeq h\mathfrak{g}$ restricted to the Lie algebra
quotient spaces $h\mathfrak{p=}h\mathfrak{g/}h\mathfrak{h,}$ with $%
T_{x}hH\simeq h\mathfrak{h,}$ where $h\mathfrak{g=}h\mathfrak{h}\oplus h%
\mathfrak{p}$ is stated such that there is an involutive automorphism of $hG$
under $hH$ is fixed, i.e. $[h\mathfrak{h,}h\mathfrak{p]}\subseteq $ $h%
\mathfrak{p}$ and $[h\mathfrak{p,}h\mathfrak{p]}\subseteq h\mathfrak{h.}$ We
can also define the group spaces and related inner products and\ Lie
algebras,%
\begin{eqnarray}
\mbox{for\ }vg &=&\{\mathring{h}_{ab}\},\;<\cdot ,\cdot
>_{v},\;T_{y}vG\simeq v\mathfrak{g,\;}v\mathfrak{p=}v\mathfrak{g/}v\mathfrak{%
h,}\mbox{ with }  \notag \\
T_{y}vH &\simeq &v\mathfrak{h,}v\mathfrak{g=}v\mathfrak{h}\oplus v\mathfrak{%
p,}\mbox{where }\mathfrak{\;}[v\mathfrak{h,}v\mathfrak{p]}\subseteq v%
\mathfrak{p,\;}[v\mathfrak{p,}v\mathfrak{p]}\subseteq v\mathfrak{h;}  \notag
\\
&&  \label{algstr} \\
\mbox{for\ }\mathbf{g} &=&\{\mathring{g}_{\alpha \beta }\},\;<\cdot ,\cdot
>_{\mathbf{g}},\;T_{(x,y)}G\simeq \mathfrak{g,\;p=g/h,}\mbox{ with }  \notag
\\
T_{(x,y)}H &\simeq &\mathfrak{h,g=h}\oplus \mathfrak{p,}\mbox{where }%
\mathfrak{\;}[\mathfrak{h,p]}\subseteq \mathfrak{p,\;}[\mathfrak{p,p]}%
\subseteq \mathfrak{h.}  \notag
\end{eqnarray}%
Any metric structure with constant coefficients on $V=G/SO(n+m)$ can be
parametrized in the form%
\begin{equation*}
\mathring{g}=\mathring{g}_{\alpha \beta }du^{\alpha }\otimes du^{\beta },
\end{equation*}%
where $u^{\alpha }$ are local coordinates and
\begin{equation}
\mathring{g}_{\alpha \beta }=\left[
\begin{array}{cc}
\mathring{g}_{ij}+\mathring{N}_{i}^{a}\mathring{N}_{j}^{b}\mathring{h}_{ab}
& \mathring{N}_{j}^{e}\mathring{h}_{ae} \\
\mathring{N}_{i}^{e}\mathring{h}_{be} & \mathring{h}_{ab}%
\end{array}%
\right] .  \label{constans}
\end{equation}%
The constant (trivial) N--connection coefficients in (\ref{constans}) are
computed $\mathring{N}_{j}^{e}=\mathring{h}^{eb}\mathring{g}_{jb}$ for any
given sets $\mathring{h}^{eb}$ and $\mathring{g}_{jb},$ i.e. from the
inverse metrics coefficients defined respectively on $hG=SO(n+1)$ and by
off--blocks $(n\times n)$-- and $(m\times m)$--terms of the metric $%
\mathring{g}_{\alpha \beta }.$ This way, we can define an equivalent
d--metric structure of type (\ref{gpsm})
\begin{eqnarray}
\mathbf{\mathring{g}} &=&\ \mathring{g}_{ij}\ e^{i}\otimes e^{j}+\ \mathring{%
h}_{ab}\ \mathbf{\mathring{e}}^{a}\otimes \mathbf{\mathring{e}}^{b},
\label{m1const} \\
e^{i} &=&dx^{i},\ \;\mathbf{\mathring{e}}^{a}=dy^{a}+\mathring{N}%
_{i}^{a}dx^{i}  \notag
\end{eqnarray}%
defining a trivial $(n+m)$--splitting $\mathbf{\mathring{g}=}\mathring{g}%
\mathbf{\oplus _{\mathring{N}}}\mathring{h}\mathbf{\ }$because all
nonholonomy coefficients $\mathring{W}_{\alpha \beta }^{\gamma }$ and
N--connection curvature coefficients $\mathring{\Omega}_{ij}^{a}$ are zero.

It is possible to consider any covariant coordinate transforms of (\ref%
{m1const}) preserving the\ $(n+m)$--splitting resulting in $w_{\alpha \beta
}^{\gamma }=0,$ see (\ref{anhrel}) and $\Omega _{ij}^{a}=0$ (\ref{ncurv}).
Such trivial parametrizations define algebraic classifications of \
symmetric Riemannian spaces of dimension $n+m$ with constant matrix
curvature admitting splitting (by certain algebraic constraints) into
symmetric Riemannian subspaces of dimension $n$ and $m,$ also both with
constant matrix curvature. This way, we get the simplest example of
nonholonomic Riemannian space of type $\mathbf{\mathring{V}}=[hG=SO(n+1),$ $%
vG=SO(m+1),\;\mathring{N}_{i}^{e}]$ possessing a Lie d--algebra symmetry $%
\mathfrak{so}_{\mathring{N}}(n+m)\doteqdot \mathfrak{so}(n)\oplus \mathfrak{%
so}(m).$

We can generalize the constructions if we consider nonholonomic
distributions on $V=G/SO(n+m)$ defined locally by arbitrary N--connection
coefficients $N_{i}^{a}(x,y),$ with nonvanishing $w_{\alpha \beta }^{\gamma
} $ and $\Omega _{ij}^{a}$ but with constant d--metric coefficients when
\begin{eqnarray}
\mathbf{\mathring{g}} &\mathbf{=g}&=\ \mathring{g}_{ij}\ e^{i}\otimes
e^{j}+\ \mathring{h}_{ab}\ \mathbf{e}^{a}\otimes \mathbf{e}^{b},  \label{m1b}
\\
e^{i} &=&dx^{i},\ \mathbf{e}^{a}=dy^{a}+N_{i}^{a}(x,y)dx^{i}.  \notag
\end{eqnarray}%
This metric is equivalent to a d--metric $\mathbf{g}_{\alpha ^{\prime }\beta
^{\prime }}=[g_{i^{\prime }j^{\prime }},h_{a^{\prime }b^{\prime }}]$ (\ref%
{gpsm}) with constant coefficients induced by the corresponding Lie
d--algebra structure\newline
$\mathfrak{so}_{\mathring{N}}(n+m).$ Such spaces transform into nonholonomic
manifolds $\mathbf{\mathring{V}}_{\mathbf{N}}=[hG=SO(n+1),$ $%
vG=SO(m+1),\;N_{i}^{e}]$ \ with nontrivial N--connection curvature and
induced d--torsion coefficients of the d--connection (\ref{dcon}). One has
zero curvature for this d--connection (in general, such spaces are curved
ones with generic off--diagonal metric (\ref{m1b}) and nonzero curvature
tensor for the Levi--Civita connection). So, such nonholonomic manifolds
posses the same group and algebraic structures of couples of symmetric
Riemannian spaces of dimension $n$ and $m$ but nonholonomically soldered to
the symmetric Riemannian space of dimension $n+m.$ With respect to
N--adapted orthonormal bases, with distinguished h-- and v--subspaces, we
obtain the same inner products and group and Lie algebra spaces as in (\ref%
{algstr}).

The bi--Hamiltonian and solitonic constructions are based on an extrinsic
approach soldering the Riemannian symmetric--space geometry to the Klein
geometry \cite{sharpe}. For the N--anholonomic spaces of dimension $n+m,$
with a constant d--curvature, similar constructions hold true but we have to
adapt them to the N--connection structure. In Ref. \cite{vcfl2}, we proved
that any (pseudo) Riemannian metric $\mathbf{g}$ on $\mathbf{V}$ defines a
set of metric compatible d--connections of type
\begin{equation}
\ _{0}\widetilde{\mathbf{\Gamma }}_{\ \alpha ^{\prime }\beta ^{\prime
}}^{\gamma ^{\prime }}=\left( \widehat{L}_{j^{\prime }k^{\prime
}}^{i^{\prime }}=0,\widehat{L}_{b^{\prime }k^{\prime }}^{a^{\prime }}=\ _{0}%
\widehat{L}_{b^{\prime }k^{\prime }}^{a^{\prime }}=const,\widehat{C}%
_{j^{\prime }c^{\prime }}^{i^{\prime }}=0,\widehat{C}_{b^{\prime }c^{\prime
}}^{a^{\prime }}=0\right)  \label{ccandcon}
\end{equation}%
with respect to N--adapted frames (\ref{dder}) and (\ref{ddif}) for\ any $%
\mathbf{N}=\{N_{i^{\prime }}^{a^{\prime }}(x,y)\}$ being a nontrivial
solution of the system of equations%
\begin{equation}
2\ _{0}\widehat{L}_{b^{\prime }k^{\prime }}^{a^{\prime }}=\frac{\partial
N_{k^{\prime }}^{a^{\prime }}}{\partial y^{b^{\prime }}}-\ \mathring{h}%
^{a^{\prime }c^{\prime }}\ \mathring{h}_{d^{\prime }b^{\prime }}\frac{%
\partial N_{k^{\prime }}^{d^{\prime }}}{\partial y^{c^{\prime }}}
\label{auxf1}
\end{equation}%
for any nondegenerate constant--coefficients symmetric matrix $\mathring{h}%
_{d^{\prime }b^{\prime }}$ and its inverse $\ \mathring{h}^{a^{\prime
}c^{\prime }}.$ Here, we emphasize that the coefficients $\ _{\shortmid
}\Gamma _{\ \alpha ^{\prime }\beta ^{\prime }}^{\gamma ^{\prime }}$ of the
corresponding to $\mathbf{g}$ Levi--Civita connection $\ ^{\mathbf{g}}\nabla
$ are not constant with respect to N--adapted frames.

By straightforward computations, we get that the curvature d--tensor of a
d--connection $\ _{0}\widetilde{\mathbf{\Gamma }}_{\ \alpha ^{\prime }\beta
^{\prime }}^{\gamma ^{\prime }}$ (\ref{ccandcon}) defined by a metric $%
\mathbf{g}$ has constant coefficients
\begin{eqnarray*}
\ _{0}\widetilde{\mathbf{R}}_{\ \beta ^{\prime }\gamma ^{\prime }\delta
^{\prime }}^{\alpha ^{\prime }} &=&(\ _{0}\widetilde{R}_{~h^{\prime
}j^{\prime }k^{\prime }}^{i^{\prime }}=0,\ _{0}\widetilde{R}_{~b^{\prime
}j^{\prime }k^{\prime }}^{a^{\prime }}=\ _{0}\widehat{L}_{\ b^{\prime
}j^{\prime }}^{c^{\prime }}\ _{0}\widehat{L}_{\ c^{\prime }k^{\prime
}}^{a^{\prime }}-\ _{0}\widehat{L}_{\ b^{\prime }k^{\prime }}^{c^{\prime }}\
_{0}\widehat{L}_{\ c^{\prime }j^{\prime }}^{a^{\prime }}= \\
&&cons,\ _{0}\widetilde{P}_{~h^{\prime }j^{\prime }a^{\prime }}^{i^{\prime
}}=0,\ _{0}\widetilde{P}_{~b^{\prime }j^{\prime }a^{\prime }}^{c^{\prime
}}=0,\ _{0}\widetilde{S}_{~j^{\prime }b^{\prime }c^{\prime }}^{i^{\prime
}}=0,\ _{0}\widetilde{S}_{~b^{\prime }d^{\prime }c^{\prime }}^{a^{\prime
}}=0)
\end{eqnarray*}
with respect to N--adapted frames $\mathbf{e}_{\alpha ^{\prime }}=[\mathbf{e}%
_{i^{\prime }},e_{a^{\prime }}]$ and $\mathbf{e}^{\alpha ^{\prime
}}=[e^{i^{\prime }},\mathbf{e}^{a^{\prime }}]$ when $N_{k^{\prime
}}^{d^{\prime }}$ are subjected to the conditions (\ref{auxf1}). Using a
deformation relation of type (\ref{deflc}), we can compute the corresponding
Ricci tensor $\ _{\shortmid }R_{\ \beta \gamma \delta }^{\alpha }$ for the
Levi--Civita connection $\ ^{\mathbf{g}}\nabla ,$ which is a general one
with 'non-constant' coefficients with respect to any local frames.

A d--connection $\ _{0}\widetilde{\mathbf{\Gamma }}_{\ \alpha ^{\prime
}\beta ^{\prime }}^{\gamma ^{\prime }}$ (\ref{ccandcon}) has constant scalar
curvature,
\begin{equation*}
\ _{0}^{\sim }\overleftrightarrow{\mathbf{R}}\doteqdot \ _{0}\mathbf{g}%
^{\alpha ^{\prime }\beta ^{\prime }}\ _{0}\widetilde{\mathbf{R}}_{\alpha
^{\prime }\beta ^{\prime }}=\ \mathring{g}^{i^{\prime }j^{\prime }}\ _{0}%
\widetilde{R}_{i^{\prime }j^{\prime }}+\ \mathring{h}^{a^{\prime }b^{\prime
}}\ _{0}\widetilde{S}_{a^{\prime }b^{\prime }}=\ _{0}^{\sim }\overrightarrow{%
R}+\ _{0}^{\sim }\overleftarrow{S}=const.
\end{equation*}%
Nevertheless, the scalar curvature $\ _{\nabla }R$ of $\ ^{\mathbf{g}}\nabla
,$ for the same metric, is not constant.

The constructions with different types of metric compatible connections
generated by a metric structure are summarized in Table 1.
\begin{table}[tbp]
\caption{Some metric connections generated by a d--metric $\mathbf{g}=\{g_{%
\protect\alpha \protect\beta} \}$ }
\label{tab1}\vskip5pt {\footnotesize \textrm{%
\begin{tabular*}{\textwidth}{@{}l@{\extracolsep{0pt plus11pt}}l|@{\extracolsep{0pt plus11pt}}l|@{\extracolsep{0pt plus11pt}}l|}
\hline\hline
\vline \qquad Geometric & \vline \qquad Levi--Civita & \ normal d--connection
& \ constant coefficients \\
\vline \qquad objects for: & \vline \qquad connection &  & \qquad
d--connection \\ \hline\hline
\vline Co-frames & \vline \ $e_{\ }^{\beta }= A_{\ \underline{\beta}}^{\beta
}(u)du^{\underline{\beta}}$ & \ $\mathbf{e}^{\alpha}= [e^i=dx^i, $ & \ $%
\mathbf{e}^{{\alpha}^{\prime}}= [e^{i^{\prime}}=dx^{i^{\prime}}, $ \\
\vline & \vline & \qquad$\mathbf{e}^a=dy^a-N^a_j dx^j]$ & \quad ${\mathbf{e}}%
^{a^{\prime}}=dy^{a^{\prime}}- N^{a^{\prime}}_{j^{\prime}} dx^{j^{\prime}}]$
\\ \hline
\vline Metric decomp. & \vline \ $g_{\alpha \beta}= A^{\ \underline{\alpha}%
}_{\alpha } A^{\ \underline{\beta}}_{\beta }g_{\underline{\alpha} \underline{%
\beta}}$ & \quad $\mathbf{g}_{\alpha \beta}= [g_{ij},h_{ab}],\ $ \  & $\ _{0}%
\mathbf{g}_{\alpha ^{\prime }\beta ^{\prime }}= [\ \mathring{g}_{i ^{\prime
}j ^{\prime }},\ \mathring{h}_{a ^{\prime }b ^{\prime }}],\ $ \\
\vline & \vline & \quad $\mathbf{g}=g_{ij}\ e^{i}\otimes e^{j}$ & \quad $%
\mathbf{g}=\mathring{g}_{i ^\prime j ^\prime}\ e^{i ^{\prime }}\otimes e^{j
^{\prime }}$ \\
\vline & \vline & \qquad $+h_{ab}\ \mathbf{e}^{a}\otimes \mathbf{e}^{b}$ &
\qquad $+ \mathring{h}_{a ^\prime b ^\prime}\ \mathbf{e}^{a ^\prime}\otimes
\mathbf{e}^{b ^\prime}$ \\
\vline & \vline &  & \ $g_{i^{\prime }j^{\prime }}=A_{\ i^{\prime }}^{i}A_{\
j^{\prime}}^{j}g_{ij},$ \\
\vline & \vline &  & \ $h_{a^{\prime }b^{\prime }}=A_{\ a^{\prime }}^{a}A_{\
b^{\prime }}^{b}h_{ab}$ \\ \hline
\vline Connections & \vline \qquad $_{\shortmid }\Gamma _{\ \alpha \beta
}^{\gamma }$ & $\ _{\shortmid }\Gamma _{\ \alpha \beta }^{\gamma }=\widehat{%
\mathbf{\Gamma }}_{\ \alpha \beta }^{\gamma }+\ _{\shortmid }Z_{\ \alpha
\beta }^{\gamma }$ & $\ _{0}\widetilde{\mathbf{\Gamma }}_{\ \alpha ^{\prime
}\beta ^{\prime}}^{\gamma ^{\prime }} =( \widehat{L}_{j^{\prime
}k^{\prime}}^{i^{\prime }}=0,$ \\
\vline and distorsions & \vline &  & \ $\widehat{L}_{b^{\prime }k^{\prime
}}^{a^{\prime }}=\ _{0}\widehat{L}_{b^{\prime }k^{\prime }}^{a^{\prime
}}=const.,$ \\
\vline & \vline &  & \qquad $\widehat{C}_{j^{\prime }c^{\prime
}}^{i^{\prime}}=0, \widehat{C}_{b^{\prime }c^{\prime }}^{a^{\prime }}=0)$ \\
\hline
\vline Riemannian & \vline \qquad $_{\shortmid }R_{~\beta \gamma \delta
}^{\alpha }$ & $\qquad \widehat{\mathbf{R}}_{~\beta \gamma \delta }^{\alpha
} $ & $\ _{0}\widetilde{\mathbf{R}}_{\ \beta ^{\prime }\gamma ^{\prime
}\delta ^{\prime }}^{\alpha ^{\prime }} =(0,\ _{0}\widetilde{R}%
_{~b^{\prime}j^{\prime}k^{\prime}}^{a^{\prime}}$ \\
\vline (d--)tensors & \vline &  & $\qquad =const.,0,0,0,0)$ \\ \hline\hline
\end{tabular*}
}}
\end{table}

We conclude that the algebraic structure of nonholonomic spaces enabled with
N--connection structure is defined by a conventional splitting of dimensions
with certain holonomic and nonholonomic variables (defining a distribution
of horizontal and vertical subspaces). Such subspaces are modelled locally
as Riemannian symmetric manifolds and their properties are exhausted by the
geometry of distinguished Lie groups $\mathbf{G}=GO(n)\oplus $ $GO(m)$ and $%
\mathbf{G}=SU(n)\oplus $ $SU(m)$ and the geometry of N--connections on a
conventional vector bundle with base manifold $\ M,$ $\dim M=n,$ and typical
fiber $F,$ $\dim F=m.$ For constructions related to Einstein gravity, we
have to consider $n=2$ and $m=2.$ This can be formulated equivalently in
terms of geometric objects on couples of Klein spaces. The bi--Hamiltonian
and related solitonic (of type mKdV and SG) hierarchies are generated
naturally by wave map equations and recursion operators associated to the
horizontal and vertical flows of curves on such spaces \cite%
{vcfl1,vcfl2,vcflrf3,vcfanco4}.

The approach allowed us to elaborate a ''solitonic'' formalism when the
geometry of (semi) Riemannian / Einstein manifolds is encoded into
nonholonomic hierarchies of bi--Hamiltonian structures and related solitonic
equations derived for curve flows on spaces with conventional splitting of
dimensions. The same distinguished group (d--group) formalism may be applied
for quantum string models and nonholonomic A--branes for the almost K\"{a}%
hler model of Einstein gravity.

\subsection{Nonholonomic deformations and quantization of distinguished
Chern--Simons gravity theories}

Let us consider a metric field $\mathbf{g}$ which is a solution of usual
Einstein equations (\ref{einsteq}). For different purposes, we can work
equivalently with any linear connection $\ ^{\mathbf{g}}\nabla =\{\
_{\shortmid }^{\mathbf{g}}\Gamma _{\ \beta \gamma }^{\alpha }\},\widehat{%
\mathbf{D}}=\{\ \widehat{\mathbf{\Gamma }}_{\ \alpha \beta }^{\gamma }\},$
or $\ _{0}\widetilde{\mathbf{D}}=\{\ _{0}\widetilde{\mathbf{\Gamma }}_{\
\alpha \beta }^{\gamma }\},$ for $\mathbf{g}$ and/or $\ ^{\mathbf{g}}\mathbf{%
\theta (X,Y)}\doteqdot \mathbf{g}\left( \mathbf{J}\mathbf{X,Y}\right) .$ For
any $\ ^{\mathbf{g}}\mathbf{\theta =(}\ ^{h}\theta ,\ ^{v}\theta \mathbf{),}$
we can associate two symplectic forms related formally to a Chern--Simons
theory with a compact gauge d--group $\mathbf{G}=GO(2)\oplus $ $GO(2),$ or $%
\mathbf{G}=SU(2)\oplus $ $SU(2)$ (for simplicity, hereafter we shall
consider the case of unitary d--groups, one for the h--part, $\ ^{h}G=SU(2),$
and another for the v--part, $\ ^{v}G=SU(2)).$ We chose two two--manifolds
without boundary, denoted $\ ^{h}C$ and $\ ^{v}C$ and consider $\ ^{\mathbf{G%
}}\mathbf{V}=(\ ^{h}V,\ ^{v}V)$ to be defined by a couple of moduli spaces
of homomorphisms (up to conjugation) from $\pi _{1}(\ ^{h}C)$ into $\ ^{h}G$
and, respectively, $\pi _{1}(\ ^{v}C)$ into $\ ^{v}G,$ of given topological
types. We can consider the same local coordinate parametrization for $\ ^{%
\mathbf{G}}\mathbf{V}$ and open regions of a nonholonomic spacetime $\mathbf{%
V.}$ The further geometric constructions are related to a symplectic
d--structure on the infinite--dimensional linear d--spaces $\ ^{\mathbf{G}}%
\mathcal{A}=(\ ^{h}\mathcal{A},\ ^{v}\mathcal{A})$ as all couples of
d--connections on a distinguished $\mathbf{G}$--bundle $\mathbf{E}%
\rightarrow \mathbf{C},$ for $\mathbf{C}=\ ^{h}C\oplus \ ^{v}C.$

\begin{figure}[tbph]
\begin{center}
\begin{picture}(360,350)
\thinlines
\put(190,340){\oval(160,35)}
\put(114,338){\makebox{
$\begin{array}{c}
 \mbox{\quad Levi--Civita variables: }\\
 \mbox{\quad $(\mathbf{g},\ ^{\mathbf{g}}\nabla )= ( g_{\mu \nu }, \ _{\shortmid }^{\mathbf{g}}%
\Gamma _{\ \beta \gamma }^{\alpha } )$}
 \end{array}$
}}
\put(20,290){\makebox{\it 2+2 nonholonomic splitting; }}

\put(20,280){\makebox{\it generating function $L(x,y)$}}
\put(240,290){\makebox{\it nonholonomic N--adapted}}
\put(250,280){\makebox{\it frame transforms }}
\put(20,210){\framebox(140,40){
$\begin{array}{c}
 \mbox{almost K\"{a}hler variables: }\\
 \mbox{$(\ ^{\mathbf{g}}\mathbf{\theta} ,\ \widehat{\mathbf{D}},\ ^L\widehat{\mathbf{N}}) = (\mathbf{\theta}_{\alpha \beta}, \widehat{\mathbf{\Gamma }}_{\ \alpha \beta
}^{\gamma } ) $}
 \end{array}$
}}
\put(215,210){\framebox(160,40){
$\begin{array}{c}
 \mbox{constant coefficient variables: }\\
 \mbox{$(\ ^\circ \mathbf{g}_{\alpha \beta} ,\ ^\circ \mathbf{D},\ {\mathbf{N}}) $}
 \end{array}$
}}
\put(20,130){\framebox(140,40)
{$\begin{array}{c}
 \mbox{metricity: }\widehat{\mathbf{D}}{\mathbf{g}}=0,\\
\ _{\shortmid }^{\mathbf{g}}\Gamma _{\ \beta \gamma }^{\alpha} = \widehat{\mathbf{\Gamma }}_{\ \beta \gamma }^{\alpha}
+\ _{\shortmid }^{\mathbf{g}}Z_{\ \beta \gamma }^{\alpha}
 \end{array}$
 }}
\put(215,130){\framebox(160,40)
{$\begin{array}{c}
 \mbox{metricity: } \ ^\circ \mathbf{D}\ ^\circ \mathbf{g}=0,\\
\ _{\shortmid }^{\mathbf{g}}\Gamma _{\ \beta \gamma }^{\alpha} =  \ ^\circ \mathbf{\Gamma}_{\ \beta \gamma }^{\alpha}
+\ _{\shortmid }^{\circ}Z_{\ \beta \gamma }^{\alpha}
 \end{array}$
 }}
\put(110,70){\dashbox(160,40){
 \mbox{Classical Einstein Spaces}}
}

\put(288,210){\shortstack[r]{ \vector(0,-1){40}}}
\put(92,210){\shortstack[r]{ \vector(0,-1){40}}}

\put(288,127){\shortstack[r]{ \vector(-1,-1){15}}}
\put(92,127){\shortstack[r]{ \vector(1,-1){15}}}

\put(190,68){\shortstack[r]{ \vector(0,-1){30}}}
\put(85,5){\framebox(210,30)
{ \mbox{Quantum almost K\"{a}hler Einstein Spaces}
}}
\put(190,320){\shortstack[r]{\vector(1,-1){70}}}
\put(190,320){\shortstack[r]{\vector(-1,-1){70}}}


\end{picture}
\end{center}
\caption{\textbf{The Levi--Civita, normal and constant coefficients
connections in Einstein gravity}}
\label{fig2}
\end{figure}
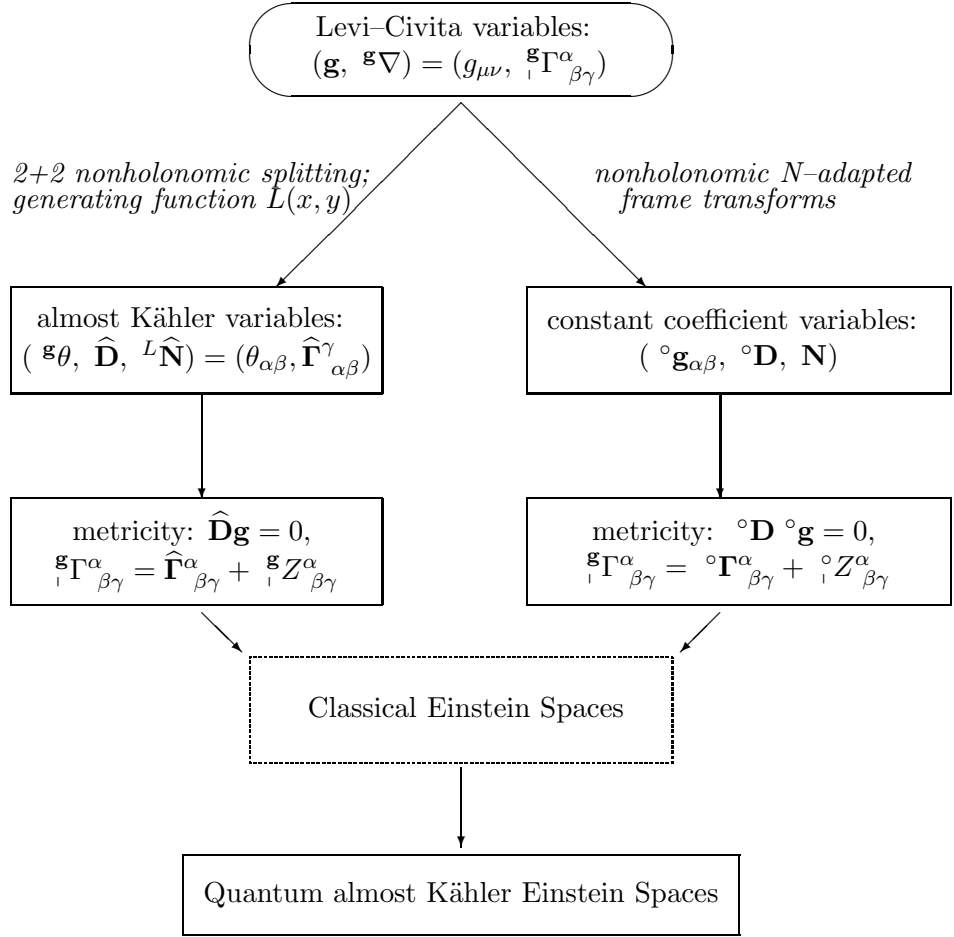

We fix such a parametrization of coefficients of a gravitation symplectic
d--form $^{\mathbf{g}}\mathbf{\theta =(}\ ^{h}\theta ,\ ^{v}\theta )$ which
in a point of $\mathbf{V}$ is proportional to the coefficients
\begin{equation}
\ _{\ast }^{h}\theta =\frac{1}{4\pi }\int_{\ ^{h}C}Tr\ \delta \ ^{h}A\wedge
\delta \ ^{h}A\mbox{\ and\ }\ _{\ast }^{v}\theta =\frac{1}{4\pi }\int_{\
^{v}C}Tr\ \delta \ ^{v}A\wedge \delta \ ^{v}A,  \label{oneforms}
\end{equation}%
when, in brief, in ''boldfaced'' form, $\ _{\ast }^{\mathbf{g}}\mathbf{%
\theta =}$ $\frac{1}{4\pi }\int_{\ \mathbf{C}}\mathbf{Tr}\ \delta \mathbf{A}%
\wedge \delta \mathbf{A.}$ The trace symbol $Tr$ is considered respectively
for the $h$-- and $v$--forms, as invariant ones on the Lie algebras $\ ^{h}%
\mathfrak{g},$ of $\ ^{h}G,$ and $\ ^{v}\mathfrak{g},$ of $\ ^{v}G,$ in our
case, in the $2$--dimensional representation. Such $\ _{\ast }^{h}\theta ,$
or $\ _{\ast }^{v}\theta ,$ are normalized to $_{\ast }\theta /2\pi $ being
the image in de Rham cohomology of a generator of $H^{2}(\ ^{h}V,\mathbb{Z}%
)\cong \mathbb{Z},$ or $H^{2}(\ ^{v}V,\mathbb{Z})\cong \mathbb{Z}.$ Here, it
should be emphasized that such local identifications of the gravitational
almost K\"{a}hler symplectic structures with couples of symplectic structure
of respective Chern--Simons theories (with $\ ^{\mathbf{G}}\mathbf{V}$
modelling the classical phase d--spaces for such models) do not impose
elaboration of classical and/or gravitational models with structures
d--groups of type $\mathbf{G.}$ We only fixed an explicit common
parametrization for a ''background'' curve flow network the chosen method of
quantization and generating gravitational solitonic hierarchies, like in %
\cite{vcfl2}. Real classical/quantum Einstein gravitational interactions can
be generated by deformations of connections $\ ^{\mathbf{g}}\nabla =\ ^{%
\mathbf{g}}D+\ ^{\mathbf{g}}Z$ (\ref{condeform}), where $\ ^{\mathbf{g}}D$
is any necessary type connection, for instance, parametrized as a gauge one
in a Chern--Simons d--model, $\mathbf{A=}(\ ^{h}A,\ ^{v}A)$ with
coefficients determined by a d--connection (\ref{dconf}), or any its
nonholonomic transform, but the related$\ ^{\mathbf{g}}Z$ is such a
distorsion tensor which nonholonomically deforms $\ ^{\mathbf{g}}D$ into $^{%
\mathbf{g}}\nabla $ defined by an Einstein solution, in the classical limit.
Any such schemes with equivalent geometric objects defined by a d--metric $%
\mathbf{g}$ but for suitable nonholonomic and topologic structures
correspond to a N--connection splitting and adapted frames as it is
described in Figure \ref{fig2}.

Next we introduce a distinguished line bundle (line d--bundle).\footnote{%
This can be constructed as in usual Chern--Simons theory \cite{ramad,bar}
but using d--groups.} Fixing two integers $\widehat{\mathbf{k}}=(\ ^{h}%
\widehat{k},\ ^{v}\widehat{k}),$ we can quantize our nonholonomic model $\ ^{%
\mathbf{G}}\mathbf{V}$ as in \cite{gukwit} but using a symplectic d--form $\
_{\ast }\mathbf{\theta =(}\ ^{h}\widehat{k}\ _{\ast }^{h}\theta ,\ ^{v}%
\widehat{k}\ _{\ast }^{v}\theta \mathbf{)}$ with a prequantum line d--bundle
$\ _{\ast }\mathcal{V}=\ _{\ast }^{\ ^{h}\widehat{k}}\mathcal{V}\oplus \
_{\ast }^{\ ^{v}\widehat{k}}\ \mathcal{V}.$ Taking $\ ^{\mathbf{G}}\mathbf{Y}
$ to be the distinguished moduli space of homomorphysms (preserving the
splitting by a prescribed N--connection structure) from $\pi _{1}(\mathbf{C}%
) $ to $\mathbf{G}_{\mathbb{C}}=\ ^{h}G_{\mathbb{C}}\oplus \ ^{v}G_{\mathbb{C%
}},$ for $\mathbf{C}=\ ^{h}C\oplus \ ^{v}C$ and $^{h}G_{\mathbb{C}}$ and $\
^{v}G_{\mathbb{C}}$ being the respective complexifications of the respective
h-- and v--groups (up to conjugation), we define $\ ^{\mathbf{G}}\mathbf{Y}$
as a natural noholonomic complex manifold.

We denote by $\mathbf{r}=(\ ^{h}r,\ ^{v}r)$ certain finite--dimensional
representations of complex Lie d--group $\ ^{h}G_{\mathbb{C}}\oplus \ ^{v}G_{%
\mathbb{C}}$ and consider two oriented closed loops $\mathbf{s}=(\ ^{h}s,\
^{v}s)$ on $\ ^{h}C\oplus \ ^{v}C$ and the holonomies of respective two flat
connections (defining a d--connection) around $\mathbf{s,}$ denoted
respectively $Hol(\mathbf{s})=Hol(\ ^{h}s)\oplus Hol(\ ^{v}s).$ This way, we
define
\begin{equation}
\mathbf{H}_{\mathbf{r}}(\mathbf{s})=Tr_{\ ^{h}r}Hol(\ ^{h}s)+Tr_{\
^{v}r}Hol(\ ^{v}s)  \label{holf}
\end{equation}%
is a holomorphic function on $\ ^{\mathbf{G}}\mathbf{Y.}$

For a gauge $\mathfrak{g}_{\mathbb{C}}$--valued d--connection $\mathbf{A=}(\
^{h}A,\ ^{v}A),$ the function (\ref{holf}) can be written using an oriented
exponential product $P$ on both $h$-- and $v$--subspaces,
\begin{equation*}
\mathbf{H}_{\mathbf{r}}(\mathbf{s})=Tr\ P\exp \left( -\oint_{\ ^{h}s}\
^{h}A\right) +Tr\ P\exp \left( -\oint_{\ ^{v}s}\ ^{v}A\right) .
\end{equation*}%
The restrictions of such holomorphic functions on $\ ^{\mathbf{G}}\mathbf{Y}$
define a dense set of functions on $^{\mathbf{G}}\mathbf{V}$ associated to $%
\mathbf{V.}$ Using nonholonomic transforms preserving the N--connection
structure, and corresponding deformations of d--connections, we can relate
such distinguished group constructions to those with a gravitationally
induced symplectic form on $\mathbf{Y.}$\footnote{%
The end of this section, we shall discuss how we can nonholonomically deform
the manifold $^{\mathbf{G}}\mathbf{Y}$ and its fundamental geometric
symplectic structures into a manifold $\mathbf{Y}$ with induced
gravitational symplectic variables.}

The $\mathfrak{g}_{\mathbb{C}}$--valued d--connection $\mathbf{A}$ also
generates a nondegenerate holomorphic distinguished 2--form $\ _{\ast }%
\mathbf{\Theta }=\left( \ _{\ast }^{h}\Theta ,\ _{\ast }^{v}\Theta \right); $
we use the complexified formulas (\ref{holf}). We can consider $^{\mathbf{G}}%
\mathbf{Y}$ as a complex symplectic manifold enabled by symplectic d--form $%
^{\mathbf{G}}\mathbf{\Theta =}\left( \ \ ^{h}\widehat{k}\ _{\ast }^{h}\Theta
,\ ^{v}\widehat{k}\ \ _{\ast }^{v}\Theta \right) ,$ with a restriction of $%
\mathbf{\Theta }$ to $^{\mathbf{G}}\mathbf{V}$ coinciding with $_{\ast }%
\mathbf{\theta .}$ This also allows us to construct a nonholonomic A--model
of $^{\mathbf{G}}\mathbf{Y}$ with symplectic structure $_{\mathbf{Y}}^{%
\mathbf{G}}\mathbf{\theta }=Im^{\mathbf{G}}\mathbf{\Theta }$ like we have
done at the beginning of section \ref{ssccnb}. Such an A--model dubs the
constructions from \cite{gukwit} (see there sections 1.3 and 2.3) and also
can be endowed with a complete hyper--K\" ahler metric (consisting from h--
and v--parts) extending its structure as a complex symplex manifold. It is a
''very good'' A--model, which allows us to pick up complex structures on $%
\mathbf{C}=\ ^{h}C\oplus \ ^{v}C$ and define complex structures on $^{%
\mathbf{G}}\mathbf{Y}$ in a natural way (requiring no structures on $\ ^{h}C$
and $\ ^{v}C$ except corresponding orientations).

There is also a natural antiholomorphic involution $\ ^{\mathbf{G}}\tau :\ ^{%
\mathbf{G}}\mathbf{Y\rightarrow \ }^{\mathbf{G}}\mathbf{Y}$ as a complex
conjugation of al monodromies (preserving h-- and v--components), were $^{%
\mathbf{G}}\mathbf{V}$ is the component of the fixed point set of $\ ^{%
\mathbf{G}}\tau $ (i.e. is is the locus in $\ ^{\mathbf{G}}\mathbf{Y}$ of
all monodromie with values in $\mathbf{G}=SU(2)\oplus $ $SU(2)).$

Having introduces two branes in the A--model of $\ ^{\mathbf{G}}\mathbf{Y,}$
we can perform the Gukov--Witten quantization of $\ ^{\mathbf{G}}\mathbf{V}.$
The first brane is a distinguished one consisting from two canonical
isotropic branes, $\ ^{\mathbf{G}}\mathcal{B}_{cc}=(\ ^{\mathbf{h}}\mathcal{B%
}_{cc},\ ^{\mathbf{v}}\mathcal{B}_{cc}),$ with curvature form ${Re}\ ^{%
\mathbf{G}}\mathbf{\Theta }$ and support by all $\ ^{\mathbf{G}}\mathbf{Y.}$
We wont to quantize the symplectic d--form $\ _{\ast }\mathbf{\theta =(}\
^{h}\widehat{k}\ _{\ast }^{h}\theta ,\ ^{v}\widehat{k}\ _{\ast }^{v}\theta )$
which is the restriction of $Re\ ^{\mathbf{G}}\mathbf{\Theta }$ to $^{%
\mathbf{G}}\mathbf{V.}$ The second brane $\ ^{\mathbf{G}}\mathcal{B}^{\prime
}$ is defined as a unique one up to an isomorphism preserving h-- and
v--splitting when $^{\mathbf{G}}\mathbf{V}$ is a simply--connected spin
manifolds; this is also a rank 1 A--brane supported on $^{\mathbf{G}}\mathbf{%
V.}$

The N--adapted diffeomorphisms of $\mathbf{C}=\ ^{h}C\oplus \ ^{v}C$ that
are continuously connected to the identity on $h$--part and identity on $v$%
--part act trivially both on the A--model of $\ ^{\mathbf{G}}\mathbf{Y}$ and
on $\ ^{\mathbf{G}}\mathbf{Y}.$ Such diffeomorphisms do not preserve hyper--K%
\"{a}hler metrics on $^{\mathbf{G}}\mathbf{Y,}$ but this is not a problem
because the A--model observables do not depend on a fixed hyper--K\"{a}hler
metric, see more details in \cite{avqg5}, on nonholonomic manifolds and
Hamilton--Cartan spaces and their deformation quantization. Following \cite%
{gukwit}, we can consider the Teichmuller space $\mathcal{T}$ of $\mathbf{C}$
when any point $\zeta \in \mathcal{T}$ determines (in a unique way, up to
isotopy) a complex structure on $\mathbf{C}$ and (as a result) a hyper--K%
\"{a}hler polarization of $\left( \ ^{\mathbf{G}}\mathbf{Y,\ }^{\mathbf{G}}%
\mathbf{V}\right) .$ The interesting thing is that the space $\ \ _{\zeta }^{%
\mathbf{G}}\mathcal{H}$ of $\left( \ \ ^{\mathbf{G}}\mathcal{B}_{cc},\ \ ^{%
\mathbf{G}}\mathcal{B}^{\prime }\right) $ strings constructed with such a
polarization does not depend on $\zeta .$ This follows from the fact that
the A--model is invariant under a local change in the hyper--K\"{a}hler
poarlization and we can also define $\ _{\zeta }^{\mathbf{G}}\mathcal{H}$ as
a typical fiber of a flat d--vector bundle $\ ^{\mathbf{G}}\mathcal{H}$ over
$\mathcal{T}.$

In general, it is a difficult task to compute exactly the (Hilbert) space $\
^{\mathbf{G}}\mathcal{H}$ for certain general nonholonomic manifolds (this
would imply the index theorem for a family of nonholonomic Dirac operators,
see \cite{vgerb1,vgerb2,vsgg}). Nevertheless, in the case relevant to
Einstein gravity and solitonic hierarchies, i.e. for $\mathbf{G}=SU(2)\oplus
$ $SU(2),$ one holds a standard (N--adapted) algebro--geometric description
of a physical Hilbert space of a nonholonomic Chern--Simons theory at levels
$\widehat{\mathbf{k}}=(\ ^{h}\widehat{k},\ ^{v}\widehat{k}),$ when $\ ^{%
\mathbf{G}}\mathcal{H}=H^{0}(\ ^{\mathbf{G}}\mathbf{V,}\ _{\ast }^{\ ^{h}%
\widehat{k}}\mathcal{V}\oplus \ _{\ast }^{\ ^{v}\widehat{k}}\ \mathcal{V}).$
Here one should be noted that in the nonholonomic A--model the (classical)
commutative algebra of holonomy functions $\mathbf{H}_{\mathbf{r}}(\mathbf{s}%
)$ (\ref{holf}) \ is nonholonomically deformed to a noncommutative
d--algebra $\mathcal{A}=(\ ^{h}\mathcal{A},\ ^{v}\mathcal{A}),$ i.e. the
space of $\left( \ ^{\mathbf{G}}\mathcal{B}_{cc},\ ^{\mathbf{G}}\mathcal{B}%
_{cc}\right) $ strings, which acts on $\ ^{\mathbf{G}}\mathcal{H}.$ When we
work with a nonholonomic Chern--Simon gauge theory, the quantization
transforms Willson loops on a distingushed $\mathbf{C}=\ ^{h}C\oplus \ ^{v}C$
into operators that acts on the quantum Hilbert space.

The Gukov--Witten quantization method is very powerful because it allows us
to consider nonholonomic deformations of geometric classical and quantum
structures and the very same d--algebra $\mathcal{A}=(\ ^{h}\mathcal{A},\
^{v}\mathcal{A})$ acts on the spaces of any nonholonomic strings for any
other choice of nonholonomic A--brane $\mathcal{B}.$ Let us explain these
new applications to gravity and nonholonomic geometries which were not
considered in \cite{gukwit}:

The coefficients of the gauge $\mathfrak{g}_{\mathbb{C}}$--valued
d--connection $\mathbf{A=}(\ ^{h}A,\ ^{v}A)$ used for constructing our
nonholonomic Chern--Simons theory can be identified with the coefficients of
a d--connection of type $\widehat{\mathbf{\Gamma }}_{j}^{i}=\widehat{L}%
_{jk}^{i}e^{k}+\widehat{C}_{jk}^{i}\mathbf{e}^{k}$ (\ref{dconf}), but with
constant d--connection coefficients, i.e. $\ _{0}\mathbf{\tilde{\Gamma}}_{\
\alpha ^{\prime }\beta ^{\prime }}^{\gamma ^{\prime }}$ (\ref{ccandcon}),
when via corresponding distorsion tensor, see Figure \ref{fig2}, $\ _{0}%
\mathbf{\tilde{\Gamma}}_{j}^{i}=\ _{0}\tilde{L}_{jk}^{i}e^{k}$ with $\
^{h}A_{jk}^{i}=\ _{0}\tilde{L}_{jk}^{i}$ and $\ ^{v}A=0$ (but this may be
for an explicit local parametrization, in general $^{v}A$ is not zero).
Further nonholonomic transforms from $\ _{0}\mathbf{\tilde{\Gamma}}$ to $%
\widehat{\mathbf{\Gamma }}$ change the nonholonomic structure of geometric
objects but does not affect the defined above hyper--K\"{a}hler
polarization. This means that the former $\mathfrak{g}_{\mathbb{C}}$--valued
d--connection structure $\mathbf{A=}(\ ^{h}A,\ ^{v}A)$ relevant to a chosen
parametrization of curve flows, for spaces $\left( \ ^{\mathbf{G}}\mathbf{%
Y,\ }^{\mathbf{G}}\mathbf{V}\right) ,$ deforms nonholonomically, but
equivalently, into a couple of nonholonomic manifolds $\left( \ ^{\mathbf{g}}%
\mathbf{Y,\ }^{\mathbf{g}}\mathbf{V}\right) $ if the metric structure of
d--group $\mathbf{G}=SU(2)\oplus $ $SU(2)$ maps nonholonomically into a
d--metric $\mathbf{g}$ (\ref{gpsm}). For such constructions, the
corresponding (classical) Levi--Civita connection $\ _{\shortmid }^{g}\Gamma
$ is constrained to be a solution of the Einstein equations (\ref{einsteq}).
Using vierbein transforms of type (\ref{algeq}) relating $(\ ^{L}\mathbf{g,}%
\ ^{L}\mathbf{N),}$ see (\ref{lfsm}) and (\ref{defcncl}), to an ''Einstein
solution'' $(\mathbf{g,}\ \mathbf{N),}$ and similar transforms to $(\mathbf{%
\mathring{g},}\ \mathbf{N)}$ (\ref{m1b}), we nonholonomically deform the
space of $\left( \ ^{\mathbf{G}}\mathcal{B}_{cc},\ ^{\mathbf{G}}\mathcal{B}%
_{cc}\right) $ strings into that of $\left( \ ^{\mathbf{g}}\mathcal{B}%
_{cc},\ ^{\mathbf{g}}\mathcal{B}_{cc}\right) $ strings. This obviously
result in equivalent transforms of $\ ^{\mathbf{G}}\mathcal{H}=H^{0}(\ ^{%
\mathbf{G}}\mathbf{V,}\ _{\ast }^{\ ^{h}\widehat{k}}\mathcal{V}\oplus \
_{\ast }^{\ ^{v}\widehat{k}}\ \mathcal{V})$ into the Hilber space $\ ^{%
\mathbf{g}}\mathcal{H}=\oplus _{\underline{j}=0}^{\dim _{\mathbb{C}}\mathbf{V%
}}H^{\underline{j}}(\mathbf{V,}\mathcal{K}^{1/2}\otimes \ ^{\mathbf{g}}%
\mathcal{V})$ (\ref{hermspace}) which is a good approximation for both
holonomic and nonholonomic quantum Einstein spaces.

Finally, we conclude that crucial for such a quantization with nonholonomic
A--branes and strings relevant to Einstein gravity are the constructions
when we define the almost K\"{a}hler variables in general relativity, in
Section \ref{sakv}, and coordinate parametrizations for almost K\"{a}hler
gravitational A--models, in Section \ref{sskgm}. The associated
constructions with a nonholonomic Chern--Simons theory for a gauge d--group $%
\mathbf{G}=SU(2)\oplus $ $SU(2)$ are also important because they allow us to
apply both a computation techniques formally elaborated for topological
gauge models and relate the constructions to further developments for
quantum curve flows, nonholonomic bi--Hamilton structures and derived
solitonic hierarchies.

\section{Final Remarks}

In the present paper we have applied the Gukov--Witten formalism \cite%
{gukwit} to quantize the almost K\"{a}hler model of Einstein gravity. We
have used some our former results on deformation and loop quantization of
gravity following ideas and methods from the geometry of nonholonomic
manifolds and (non) commutative spaces enabled with nonholonomic
distributions and associated nonlinear connections structures \cite%
{vpla,vfqlf,vloopdq,vegnakglfdq}. It was shown that the approach with
nonholonomic A--branes endowed with geometric structures induced by (pseudo)
Riemannian metrics serves to quantize standard gravity theories\footnote{%
and other more general nonholonomic geometric and physical models, for
instance, various types Finsler--Lagrange and Hamilton--Cartan spaces etc}
and redefine previous geometric results on the language of string  theories
and branes subjected to different types of nonholonomic constrains.

The A--model approach to quantization \cite{gukwit,kapust1,kapust2,bar}
seems to be efficient for elaborating quantum versions of (non) commutative
gauge models of gravity \cite{vncgr1,vncgr4}, nonholonomic Clifford--Lie
algebroid systems \cite{vcfalgebr,vvicol}, gerbes and Clifford modules \cite%
{vgerb1,vgerb2} etc when a synthesis with the methods of geometric \cite%
{konst,sour,wood,sniat} and deformation \cite{fedos,konts,karabeg1}
quantization is considered. Further perspectives are related to nonholonomic
Ricci flows and almost K\" ahler models of spaces with symmetric and
nonsymmetric metrics \cite{vnsak,vcflrf3}.

It was shown that deformation quantization of the relativistic particles
gives the same results as the canonical quantization and path integral
methods and the direction was developed for systems with second class
constraints. On such results, we cite a series of works on commutative and
noncommutative physical models of particles and strings \cite%
{sggc,gcppt,antons,lmart,krivor,derigl,hkm,castro1,castro2} and emphasize
that the Stratonovich--Weyl quantizer, Weyl correspondence, Moyal product
and the Wigner function were obtained for all the analyzed systems which
allows a straightforward generalization to nonholonomic spaces and related
models of gravity, gauge and spinor interactions and strings \cite%
{vpla,vfqlf,vncgr1,vncgr4,vstr1,vstr2}. Introducing almost K\" ahler
variables for gravity theories, such constructions and generalizations can
be deformed nonholonomically to relate (for certain well defined limits) the
Gukov--Witten quantization to deformation and geometric quantization, loop
configurations and noncommutative geometry.

One of the main results of this work is that we have shown that it is
possible to construct in explicit form a Hilbert space with a Hermitian
inner product in quantum gravity on certain couples of strings for the
almost K\" ahler model of Einstein gravity. This is completely different
from the loop quantum gravity philosophy and methods (see critical remarks,
discussions and references in \cite{nicolai,thiemann,vloopdq}) when the
background field method is not legitimate for non--perturbative
considerations.  Working with almost K\" ahler variables, it is obvious how
the constructions for deformation quantization of gravity, loop quantum
gravity (in our case with nonholonomic Ashtekar variables) and other
directions can be related to be equivalent for certain quantum and/or
classical configurations. Here, it should be noted that different methods of
quantization of nonlinear field/string theories, in general, result in very
different quantum theories.

Nevertheless, we have not analyzed the Gukov--Witten method and the almost
K\" ahler approach to gravity (relevant also to the proposed Fedosov
quantization of the Einstein theory and Lagrange--Finsler systems) in
connection to one-- and two--loops computations in perturbative gravity \cite%
{goroff,vanven}. We also have not concerned the problem of
non--renormalizability of Einstein gravity (see recent reviews of results in
Refs. \cite{bern,gomiswein}) and have not discussed the ideas on a possible
asymptotic safety scenario in quantum gravity \cite{niedermaier,lausch} in
connection to the A--model formalism and deformation quantization.

Our approach with compatible multi--connections defined by a metric
structure (in particular, by a solution of the Einstein equation) allows us
to put the above mentioned problems of non--renormalizability and safety of
gravitational interactions in a different manner, when the background
constructions are re--defined for an alternative distinguished connection
for which the one--, two-- and certain higher order loops contributions can
be formally renormalized (such constructions with necessary types of
background distinguished connections are under elaboration). But even in
such cases, the dimension of gravitational constant states explicitly that a
standard renormalization similar to gauge models is not possible for the
Einstein gravity.

Perhaps, a variant of modified by nonholonomic distorsions of gravitational
connections resulting in a gauge like model of gravity with an additional
effective constant may present interest for applications in modern cosmology
and high energy physics. To work with almost K\" ahler variables and using
the constructions for Hilbert spaces and nonholonomic methods developed in
this paper is to provide a beginning for future investigations on effective
and modified perturbative gravity models and more general non--perturbative
nonholonomic quantum geometries.

The extension of the nonlinear connection formalism to methods of
quantization and application to more general supergravity and superstring
theory \cite{strmon1,strmon2,strmon3} and spaces enabled with nonholonomic
(super) distributions \cite{vsgg,vstr2} consist a set of open problems that
may be pursued in the near future. It is interesting also to apply the
matters of almost K\" ahler variables and nonholonomic classical and quantum
interactions and geometries to more complicated second class constrained
systems as the BRST quantization in gauge and gravity theories and
Batalin--Wilkovisky quantization (see original results and reviews in \cite%
{kazin,lavr,gps}). We are going to address such topics in the future.

\vskip5pt

\textbf{Acknowledgements } The work was partially performed during a visit
at Fields Institute, Toronto.

\appendix

\setcounter{equation}{0} \renewcommand{\theequation}
{A.\arabic{equation}} \setcounter{subsection}{0}
\renewcommand{\thesubsection}
{A.\arabic{subsection}}

\section{Almost K\"{a}hler Variables in Component Form}

\label{append1}We parametrize a general (pseudo) Riemannian metric $\mathbf{g%
}$ on a spacetime $\mathbf{V}$ \ in the form: \
\begin{eqnarray}
\mathbf{g} &=&g_{i^{\prime }j^{\prime }}(u)e^{i^{\prime }}\otimes
e^{j^{\prime }}+h_{a^{\prime }b^{\prime }}(u)e^{a^{\prime }}\otimes
e^{b^{\prime }},  \label{gpsm} \\
e^{a^{\prime }} &=&\mathbf{e}^{a^{\prime }}-N_{i^{\prime }}^{a^{\prime
}}(u)e^{i^{\prime }},  \notag
\end{eqnarray}%
where the vierbein coefficients $e_{\ \alpha }^{\alpha ^{\prime }}$ of the
dual basis
\begin{equation}
e^{\alpha ^{\prime }}=(e^{i^{\prime }},e^{a^{\prime }})=e_{\ \alpha
}^{\alpha ^{\prime }}(u)du^{\alpha },  \label{dft}
\end{equation}%
define a formal $2+2$ splitting.

Let us consider any generating function $L(u)=L(x^{i},y^{a})$ on $\mathbf{V}$
(it is a formal pseudo--Lagrangian if an effective continuous mechanical
model of general relativity is elaborated, see Refs. \cite{vrflg,vsgg}) with
nondegenerate Hessian
\begin{equation}
\ ^{L}h_{ab}=\frac{1}{2}\frac{\partial ^{2}L}{\partial y^{a}\partial y^{b}},
\label{elm}
\end{equation}%
when $\det |\ ^{L}h_{ab}|\neq 0.$ We define%
\begin{eqnarray}
\ ^{L}N_{i}^{a} &=&\frac{\partial G^{a}}{\partial y^{2+i}},  \label{clnc} \\
G^{a} &=&\frac{1}{4}\ ^{L}h^{a\ 2+i}\left( \frac{\partial ^{2}L}{\partial
y^{2+i}\partial x^{k}}y^{2+k}-\frac{\partial L}{\partial x^{i}}\right) ,
\notag
\end{eqnarray}%
where $\ ^{L}h^{ab}$ is inverse to $\ ^{L}h_{ab}$ and respective
contractions of $h$-- and $v$--indices, $\ i,j,...$ and $a,b...,$ are
performed following the rule: we can write, for instance, an up $v$--index $%
a $ as $a=2+i$ and contract it with a low index $i=1,2.$ Briefly, we shall
write $y^{i}$ instead of $y^{2+i},$ or $y^{a}.$ The values (\ref{elm}) and (%
\ref{clnc}) allow us to consider
\begin{eqnarray}
^{L}\mathbf{g} &=&\ ^{L}g_{ij}dx^{i}\otimes dx^{j}+\ ^{L}h_{ab}\ ^{L}\mathbf{%
e}^{a}\otimes \ ^{L}\mathbf{e}^{b},  \label{lfsm} \\
\ ^{L}\mathbf{e}^{a} &=&dy^{a}+\ ^{L}N_{i}^{a}dx^{i},\ ^{L}g_{ij}=\
^{L}h_{2+i\ 2+j}.  \notag
\end{eqnarray}

A metric $\mathbf{g}$ (\ref{gpsm}) with coefficients $g_{\alpha ^{\prime
}\beta ^{\prime }}=[g_{i^{\prime }j^{\prime }},h_{a^{\prime }b^{\prime }}]$
computed with respect to a dual basis $e^{\alpha ^{\prime }}=(e^{i^{\prime
}},e^{a^{\prime }})$ can be related to the metric $\ ^{L}\mathbf{g}_{\alpha
\beta }=\ [\ ^{L}g_{ij},\ ^{L}h_{ab}]$ (\ref{lfsm}) with coefficients
defined with respect to a N--adapted dual basis $\ ^{L}e^{\alpha }=(dx^{i},\
^{L}\mathbf{e}^{a})$ if there are satisfied the conditions%
\begin{equation}
\mathbf{g}_{\alpha ^{\prime }\beta ^{\prime }}e_{\ \alpha }^{\alpha ^{\prime
}}e_{\ \beta }^{\beta ^{\prime }}=~^{L}\mathbf{g}_{\alpha \beta }.
\label{algeq}
\end{equation}%
Considering any given values $\mathbf{g}_{\alpha ^{\prime }\beta ^{\prime }}$
and $~^{L}\mathbf{g}_{\alpha \beta },$ we have to solve a system of
quadratic algebraic equations with unknown variables $e_{\ \alpha }^{\alpha
^{\prime }},$ see details in Ref. \cite{vloopdq}. Usually, for given values $%
[g_{i^{\prime }j^{\prime }},h_{a^{\prime }b^{\prime }},N_{i^{\prime
}}^{a^{\prime }}]$ and $\ [\ ^{L}g_{ij},\ ^{L}h_{ab},\ \ ^{L}N_{i}^{a}],$ we
can write
\begin{equation}
N_{i^{\prime }}^{a^{\prime }}=e_{i^{\prime }}^{\ i}e_{\ a}^{a^{\prime }}\ \
^{L}N_{i}^{a}  \label{defcncl}
\end{equation}%
for $e_{i^{\prime }}^{\ i}$ being inverse to $e_{\ i}^{i^{\prime }}.$

A nonlinear connection (N--connection) structure $\mathbf{N}$ on $\mathbf{V}$
can be introduced as a nonholonomic distribution (a Whitney sum)%
\begin{equation}
T\mathbf{V}=h\mathbf{V}\oplus v\mathbf{V}  \label{whitney}
\end{equation}%
into conventional horizontal (h) and vertical (v) subspaces. In local form,
a N--connection is given by its coefficients $N_{i}^{a}(u),$ when%
\begin{equation}
\mathbf{N}=N_{i}^{a}(u)dx^{i}\otimes \frac{\partial }{\partial y^{a}}.
\label{coeffnc}
\end{equation}

A N--connection on $\mathbf{V}^{n+n}$ induces a (N--adapted) frame
(vielbein) structure
\begin{equation}
\mathbf{e}_{\nu }=\left( \mathbf{e}_{i}=\frac{\partial }{\partial x^{i}}%
-N_{i}^{a}(u)\frac{\partial }{\partial y^{a}},e_{a}=\frac{\partial }{%
\partial y^{a}}\right) ,  \label{dder}
\end{equation}%
and a dual frame (coframe) structure
\begin{equation}
\mathbf{e}^{\mu }=\left( e^{i}=dx^{i},\mathbf{e}%
^{a}=dy^{a}+N_{i}^{a}(u)dx^{i}\right) .  \label{ddif}
\end{equation}%
The vielbeins (\ref{ddif}) satisfy the nonholonomy relations
\begin{equation}
\lbrack \mathbf{e}_{\alpha },\mathbf{e}_{\beta }]=\mathbf{e}_{\alpha }%
\mathbf{e}_{\beta }-\mathbf{e}_{\beta }\mathbf{e}_{\alpha }=w_{\alpha \beta
}^{\gamma }\mathbf{e}_{\gamma }  \label{anhrel}
\end{equation}%
with (antisymmetric) nontrivial anholonomy coefficients $w_{ia}^{b}=\partial
_{a}N_{i}^{b}$ and $w_{ji}^{a}=\Omega _{ij}^{a},$ where
\begin{equation}
\Omega _{ij}^{a}=\mathbf{e}_{j}\left( N_{i}^{a}\right) -\mathbf{e}_{i}\left(
N_{j}^{a}\right)  \label{ncurv}
\end{equation}%
are the coefficients of N--connection curvature (defined as the Neijenhuis
tensor on $\mathbf{V}^{n+n}).$ The particular holonomic/ integrable case is
selected by the integrability conditions $w_{\alpha \beta }^{\gamma }=0.$%
\footnote{%
we use boldface symbols for spaces (and geometric objects on such spaces)
enabled with N--connection structure}

A N--anholonomic manifold is a (nonholonomic) manifold enabled with
N--connection structure (\ref{whitney}). The geometric properties of a
N--anholonomic manifold are distinguished by some N--adapted bases (\ref%
{dder}) and (\ref{ddif}). A geometric object is N--adapted (equivalently,
distinguished), i.e. a d--object, if it can be defined by components adapted
to the splitting (\ref{whitney}) (one uses terms d--vector, d--form,
d--tensor). For instance, a d--vector $\mathbf{X}=X^{\alpha }\mathbf{e}%
_{\alpha }=X^{i}\mathbf{e}_{i}+X^{a}e_{a}$ and a one d--form $\widetilde{%
\mathbf{X}}$ (dual to $\mathbf{X}$) is $\widetilde{\mathbf{X}}=X_{\alpha }%
\mathbf{e}^{\alpha }=X_{i}e^{i}+X_{a}e^{a}.$\footnote{%
We can redefine equivalently the geometric constructions for arbitrary frame
and coordinate systems; the N--adapted constructions allow us to preserve
the conventional h-- and v--splitting.}

We introduce a linear operator $\mathbf{J}$ acting on vectors on $\mathbf{V}$
following formulas $\mathbf{J}(\mathbf{e}_{i})=-e_{2+i}$ and $\mathbf{J}%
(e_{2+i})=\mathbf{e}_{i},$ where and $\mathbf{J\circ J=-}\mathbb{I},$ for $%
\mathbb{I}$ being the unity matrix, and construct a tensor field on $\mathbf{%
V},$%
\begin{eqnarray}
\mathbf{J} &=&\mathbf{J}_{\ \beta }^{\alpha }\ e_{\alpha }\otimes e^{\beta }=%
\mathbf{J}_{\ \underline{\beta }}^{\underline{\alpha }}\ \frac{\partial }{%
\partial u^{\underline{\alpha }}}\otimes du^{\underline{\beta }}
\label{acstr} \\
&=&\mathbf{J}_{\ \beta ^{\prime }}^{\alpha ^{\prime }}\ \mathbf{e}_{\alpha
^{\prime }}\otimes \mathbf{e}^{\beta ^{\prime }}=\mathbf{-}e_{2+i}\otimes
e^{i}+\mathbf{e}_{i}\otimes \ \mathbf{e}^{2+i}  \notag \\
&=&-\frac{\partial }{\partial y^{i}}\otimes dx^{i}+\left( \frac{\partial }{%
\partial x^{i}}-\ ^{L}N_{i}^{2+j}\frac{\partial }{\partial y^{j}}\right)
\otimes \left( dy^{i}+\ ^{L}N_{k}^{2+i}dx^{k}\right) ,  \notag
\end{eqnarray}%
defining globally an almost complex structure on\ $\mathbf{V}$ completely
determined by a fixed $L(x,y).$ In this work we consider only structures $%
\mathbf{J=}\ ^{L}\mathbf{J}$ induced by a $\ ^{L}N_{i}^{2+j},$ i.e. one
should be written $\ ^{L}\mathbf{J,}$ but, for simplicity, we shall omit
left label $L,$ because the constructions hold true for any regular
generating function $L(x,y).$ Using vielbeins $\mathbf{e}_{\ \underline{%
\alpha }}^{\alpha }$ and their duals $\mathbf{e}_{\alpha \ }^{\ \underline{%
\alpha }}$, defined by $e_{\ \alpha }^{\alpha ^{\prime }}$ solving (\ref%
{algeq}), we can compute the coefficients of tensor $\mathbf{J}$ with
respect to any local basis $e_{\alpha }$ and $e^{\alpha }$ on $\mathbf{V},$ $%
\mathbf{J}_{\ \beta }^{\alpha }=\mathbf{e}_{\ \underline{\alpha }}^{\alpha }%
\mathbf{J}_{\ \underline{\beta }}^{\underline{\alpha }}\mathbf{e}_{\beta \
}^{\ \underline{\beta }}.$ In general, we can define an almost complex
structure $\mathbf{J}$ for an arbitrary N--connection $\mathbf{N,}$ stating
a nonholonomic $2+2$ splitting, by using N--adapted bases (\ref{dder}) and (%
\ref{ddif}).

The Neijenhuis tensor field for any almost complex structure $\mathbf{J}$
defined by a N--connection (equivalently, the curvature of N--connecti\-on)
is
\begin{equation}
\ ^{\mathbf{J}}\mathbf{\Omega (X,Y)\doteqdot -[X,Y]+[JX,JY]-J[JX,Y]-J[X,JY],}
\label{neijt}
\end{equation}%
for any d--vectors $\mathbf{X}$ and $\mathbf{Y.}$ With respect to N--adapted
bases (\ref{dder}) and (\ref{ddif}), a subset of the coefficients of the
Neijenhuis tensor defines the N--connection curvature,
\begin{equation}
\Omega _{ij}^{a}=\frac{\partial N_{i}^{a}}{\partial x^{j}}-\frac{\partial
N_{j}^{a}}{\partial x^{i}}+N_{i}^{b}\frac{\partial N_{j}^{a}}{\partial y^{b}}%
-N_{j}^{b}\frac{\partial N_{i}^{a}}{\partial y^{b}}.  \label{nccurv}
\end{equation}%
A N--anholonomic manifold $\mathbf{V}$ is integrable if $\Omega _{ij}^{a}=0.$
We get a complex structure if and only if both the h-- and v--distributions
are integrable, i.e. if and only if $\Omega _{ij}^{a}=0$ and $\frac{\partial
N_{j}^{a}}{\partial y^{i}}-\frac{\partial N_{i}^{a}}{\partial y^{j}}=0.$

One calls an almost symplectic structure on a manifold $\mathbf{V}$ a
nondegenerate 2--form
\begin{equation*}
\theta =\frac{1}{2}\theta _{\alpha \beta }(u)e^{\alpha }\wedge e^{\beta } =%
\frac{1}{2}\theta _{ij}(u)e^{i}\wedge e^{j}+\frac{1}{2}\theta _{ab}(u)%
\mathbf{e}^{a}\wedge \mathbf{e}^{b}.
\end{equation*}
An almost Hermitian model of a (pseudo) Riemannian spa\-ce $\mathbf{V}$
equipped with a N--connection structure $\mathbf{N}$ is defined by a triple $%
\mathbf{H}^{2+2}=(\mathbf{V},\theta ,\mathbf{J}),$ where $\mathbf{\theta
(X,Y)}\doteqdot \mathbf{g}\left( \mathbf{JX,Y}\right) $ for any $\mathbf{g}$
(\ref{gpsm}). A space $\mathbf{H}^{2+2}$ is almost K\"{a}hler, denoted $%
\mathbf{K}^{2+2},$ if and only if $d\mathbf{\theta }=0.$

For $\mathbf{g}=\ $ $^{L}\mathbf{g}$ (\ref{lfsm}) and structures $\ ^{L}%
\mathbf{N}$ (\ref{clnc}) and $\mathbf{J}$ canonically defined by $L,$ we
define $\ ^{L}\mathbf{\theta (X,Y)}\doteqdot \ ^{L}\mathbf{g}\left( \mathbf{%
JX,Y}\right) $ for any d--vectors $\mathbf{X}$ and $\mathbf{Y.}$ In local
N--adapted form form, we have
\begin{eqnarray}
\ ^{L}\mathbf{\theta } &=&\frac{1}{2}\ ^{L}\theta _{\alpha \beta
}(u)e^{\alpha }\wedge e^{\beta }=\frac{1}{2}\ ^{L}\theta _{\underline{\alpha
}\underline{\beta }}(u)du^{\underline{\alpha }}\wedge du^{\underline{\beta }}
\label{asymstr} \\
&=&\ ^{L}g_{ij}(x,y)e^{2+i}\wedge dx^{j}=\ ^{L}g_{ij}(x,y)(dy^{2+i}+\
^{L}N_{k}^{2+i}dx^{k})\wedge dx^{j}.  \notag
\end{eqnarray}%
Let us consider the form $\ ^{L}\omega =\frac{1}{2}\frac{\partial L}{%
\partial y^{i}}dx^{i}.$ A straightforward computation shows that $\ ^{L}%
\mathbf{\theta }=d\ ^{L}\omega ,$ which means that $d\ ^{L}\mathbf{\theta }%
=dd\ ^{L}\omega =0,$ i.e. the canonical effective Lagrange structures $%
\mathbf{g}=\ ^{L}\mathbf{g},\ \ ^{L}\mathbf{N}$ and $\mathbf{J}$ induce an
almost K\"{a}hler geometry. We can express the 2--form (\ref{asymstr}) as
\begin{eqnarray}
\mathbf{\theta } &=&\ ^{L}\mathbf{\theta }=\frac{1}{2}\ ^{L}\theta
_{ij}(u)e^{i}\wedge e^{j}+\frac{1}{2}\ ^{L}\theta _{ab}(u)\mathbf{e}%
^{a}\wedge \mathbf{e}^{b}  \label{canalmsf} \\
&=&g_{ij}(x,y)\left[ dy^{i}+N_{k}^{i}(x,y)dx^{k}\right] \wedge dx^{j},
\notag
\end{eqnarray}%
where the coefficients $\ ^{L}\theta _{ab}=\ ^{L}\theta _{2+i\ 2+j}$ are
equal respectively to the coefficients $\ ^{L}\theta _{ij}.$ It should be
noted that for a general 2--form $\theta $ constructed for any metric $%
\mathbf{g}$ and almost complex $\mathbf{J}$\textbf{\ }structures on $V$ one
holds $d\theta \neq 0.$ But for any $2+2$ splitting induced by an effective
Lagrange generating function, we have $d\ ^{L}\mathbf{\theta }=0.$ We have
also $d\ \mathbf{\theta }=0$ for any set of 2--form coefficients $\mathbf{%
\theta }_{\alpha ^{\prime }\beta ^{\prime }}e_{\ \alpha }^{\alpha ^{\prime
}}e_{\ \beta }^{\beta ^{\prime }}=~^{L}\mathbf{\theta }_{\alpha ^{\prime
}\beta ^{\prime }}$ (such a 2--form $\mathbf{\theta }$ will be called to be
a canonical one).

We conclude that having chosen a regular generating function $L(x,y)$ on a
(pseudo) Riemannian spacetime $\mathbf{V},$ we can always model this
spacetime equivalently as an almost K\"{a}hler manifold.

A distinguished connection (in brief, d--connection) on a spacetime $\mathbf{%
V}$,
\begin{equation}
\mathbf{D}=(hD;vD)=\{\mathbf{\Gamma }_{\beta \gamma }^{\alpha
}=(L_{jk}^{i},\ ^{v}L_{bk}^{a};C_{jc}^{i},\ ^{v}C_{bc}^{a})\},  \label{dcon}
\end{equation}%
is a linear connection which preserves under parallel transports the
distribution (\ref{whitney}). In explicit form, the coefficients\ $\mathbf{%
\Gamma }_{\beta \gamma }^{\alpha }$ are computed with respect to a
N--adapted basis (\ref{dder}) and (\ref{ddif}). A d--connection $\mathbf{D}$%
\ is metric compatible with a d--metric $\mathbf{g}$ if $\mathbf{D}_{\mathbf{%
X}}\mathbf{g}=0$ for any d--vector field $\mathbf{X.}$

If an almost symplectic structure $\theta $ is considered on a
N--anholonomic manifold, an almost symplectic d--connection $\ _{\theta }%
\mathbf{D}$ on $\mathbf{V}$ is defined by the conditions that it is
N--adapted, i.e. it is a d--connection, and $\ _{\theta }\mathbf{D}_{\mathbf{%
X}}\theta =0,$ for any d--vector $\mathbf{X.}$ From the set of metric and/or
almost symplectic compatible d--connecti\-ons on a (pseudo) Riemannian
manifold $\mathbf{V},$ we can select those which are completely defined by a
metric $\mathbf{g}=\ $ $^{L}\mathbf{g}$ (\ref{lfsm}) and an effective
Lagrange structure $L(x,y):$

There is a unique normal d--connection
\begin{eqnarray}
\ \widehat{\mathbf{D}} &=&\left\{ h\widehat{D}=(\widehat{D}_{k},^{v}\widehat{%
D}_{k}=\widehat{D}_{k});v\widehat{D}=(\widehat{D}_{c},\ ^{v}\widehat{D}_{c}=%
\widehat{D}_{c})\right\}  \label{ndc} \\
&=&\{\widehat{\mathbf{\Gamma }}_{\beta \gamma }^{\alpha }=(\widehat{L}%
_{jk}^{i},\ ^{v}\widehat{L}_{2+j\ 2+k}^{2+i}=\widehat{L}_{jk}^{i};\ \widehat{%
C}_{jc}^{i}=\ ^{v}\widehat{C}_{2+j\ c}^{2+i},\ ^{v}\widehat{C}_{bc}^{a}=%
\widehat{C}_{bc}^{a})\},  \notag
\end{eqnarray}%
which is metric compatible, $\widehat{D}_{k}\ ^{L}g_{ij}=0$ and $\widehat{D}%
_{c}\ ^{L}g_{ij}=0,$ and completely defined by a couple of h-- and
v--components $\ \widehat{\mathbf{D}}_{\alpha }=(\widehat{D}_{k},\widehat{D}%
_{c}),$ with N--adapted coefficients $\widehat{\mathbf{\Gamma }}_{\beta
\gamma }^{\alpha }=(\widehat{L}_{jk}^{i},\ ^{v}\widehat{C}_{bc}^{a}),$ where
\begin{eqnarray}
\widehat{L}_{jk}^{i} &=&\frac{1}{2}\ ^{L}g^{ih}\left( \mathbf{e}_{k}\
^{L}g_{jh}+\mathbf{e}_{j}\ ^{L}g_{hk}-\mathbf{e}_{h}\ ^{L}g_{jk}\right) ,
\label{cdcc} \\
\widehat{C}_{jk}^{i} &=&\frac{1}{2}\ ^{L}g^{ih}\left( \frac{\partial \
^{L}g_{jh}}{\partial y^{k}}+\frac{\partial \ ^{L}g_{hk}}{\partial y^{j}}-%
\frac{\partial \ ^{L}g_{jk}}{\partial y^{h}}\right) .  \notag
\end{eqnarray}%
In general, we can omit label $L$ and work with arbitrary $\mathbf{g}%
_{\alpha ^{\prime }\beta ^{\prime }}$ and $\widehat{\mathbf{\Gamma }}_{\beta
^{\prime }\gamma ^{\prime }}^{\alpha ^{\prime }}$ with the coefficients
recomputed by frame transforms (\ref{dft}).

Introducing the normal d--connection 1--form
\begin{equation}
\widehat{\mathbf{\Gamma }}_{j}^{i}=\widehat{L}_{jk}^{i}e^{k}+\widehat{C}%
_{jk}^{i}\mathbf{e}^{k},  \label{dconf}
\end{equation}
we prove that the Cartan structure equations are satisfied,%
\begin{equation}
de^{k}-e^{j}\wedge \widehat{\mathbf{\Gamma }}_{j}^{k}=-\widehat{\mathcal{T}}%
^{i},\ d\mathbf{e}^{k}-\mathbf{e}^{j}\wedge \widehat{\mathbf{\Gamma }}%
_{j}^{k}=-\ ^{v}\widehat{\mathcal{T}}^{i},  \label{cart1}
\end{equation}%
and
\begin{equation}
d\widehat{\mathbf{\Gamma }}_{j}^{i}-\widehat{\mathbf{\Gamma }}_{j}^{h}\wedge
\widehat{\mathbf{\Gamma }}_{h}^{i}=-\widehat{\mathcal{R}}_{\ j}^{i}.
\label{cart2}
\end{equation}%
The h-- and v--components of the torsion 2--form $\widehat{\mathcal{T}}%
^{\alpha }=\left( \widehat{\mathcal{T}}^{i},\ ^{v}\widehat{\mathcal{T}}%
^{i}\right) =\widehat{\mathbf{T}}_{\ \tau \beta }^{\alpha }\ \mathbf{e}%
^{\tau }\wedge \mathbf{e}^{\beta }$ from (\ref{cart1}) are computed
\begin{equation}
\widehat{\mathcal{T}}^{i}=\widehat{C}_{jk}^{i}e^{j}\wedge \mathbf{e}^{k},\
^{v}\widehat{\mathcal{T}}^{i}=\frac{1}{2}\ ^{L}\Omega _{kj}^{i}e^{k}\wedge
e^{j}+(\frac{\partial \ \ ^{L}N_{k}^{i}}{\partial y^{j}}-\widehat{L}_{\
kj}^{i})e^{k}\wedge \mathbf{e}^{j},  \label{tform}
\end{equation}%
where $\ ^{L}\Omega _{kj}^{i}$ are coefficients of the curvature of the
canonical N--connection $N_{k}^{i}$ defined by formulas similar to (\ref%
{nccurv}). The formulas (\ref{tform}) parametrize the h-- and v--components
of torsion $\widehat{\mathbf{T}}_{\beta \gamma }^{\alpha }$ in the form
\begin{equation}
\widehat{T}_{jk}^{i}=0,\widehat{T}_{jc}^{i}=\widehat{C}_{\ jc}^{i},\widehat{T%
}_{ij}^{a}=\ ^{L}\Omega _{ij}^{a},\widehat{T}_{ib}^{a}=e_{b}\left( \
^{L}N_{i}^{a}\right) -\widehat{L}_{\ bi}^{a},\widehat{T}_{bc}^{a}=0.
\label{cdtors}
\end{equation}%
It should be noted that $\widehat{\mathbf{T}}$ vanishes on h- and
v--subspaces, i.e. $\widehat{T}_{jk}^{i}=0$ and $\widehat{T}_{bc}^{a}=0,$
but certain nontrivial h--v--components induced by the nonholonomic
structure are defined canonically by $\mathbf{g}=\ ^{L}\mathbf{g}$ (\ref%
{lfsm}) and $L.$

Similar formulas holds true, for instance, for the Levi--Civita linear
connection $\bigtriangledown =\{\ _{\shortmid }\Gamma _{\beta \gamma
}^{\alpha }\}$ which is uniquely defined by a metric structure by conditions
$~\ _{\shortmid }\mathcal{T}=0$ and $\bigtriangledown \mathbf{g}=0.$ It
should be noted that this connection is not adapted to the distribution (\ref%
{whitney}) because it does not preserve under parallelism the h- and
v--distribution. Any geometric construction for the canonical d--connection $%
\widehat{\mathbf{D}}$ can be re--defined by the Levi--Civita connection by
using the formula
\begin{equation}
\ _{\shortmid }\Gamma _{\ \alpha \beta }^{\gamma }=\widehat{\mathbf{\Gamma }}%
_{\ \alpha \beta }^{\gamma }+\ _{\shortmid }Z_{\ \alpha \beta }^{\gamma },
\label{deflc}
\end{equation}%
where the both connections $\ _{\shortmid }\Gamma _{\ \alpha \beta }^{\gamma
}$ and $\widehat{\mathbf{\Gamma }}_{\ \alpha \beta }^{\gamma }$ and the
distorsion tensor $\ _{\shortmid }Z_{\ \alpha \beta }^{\gamma }$ with
N--adapted coefficients where%
\begin{eqnarray}
\ _{\shortmid }Z_{jk}^{a} &=&-C_{jb}^{i}g_{ik}h^{ab}-\frac{1}{2}\Omega
_{jk}^{a},~_{\shortmid }Z_{bk}^{i}=\frac{1}{2}\Omega
_{jk}^{c}h_{cb}g^{ji}-\Xi _{jk}^{ih}~C_{hb}^{j},  \notag \\
_{\shortmid }Z_{bk}^{a} &=&~^{+}\Xi _{cd}^{ab}~~^{\circ
}L_{bk}^{c},_{\shortmid }Z_{kb}^{i}=\frac{1}{2}\Omega
_{jk}^{a}h_{cb}g^{ji}+\Xi _{jk}^{ih}~C_{hb}^{j},\ _{\shortmid }Z_{jk}^{i}=0,
\label{deft} \\
\ _{\shortmid }Z_{jb}^{a} &=&-~^{-}\Xi _{cb}^{ad}~~^{\circ }L_{dj}^{c},\
_{\shortmid }Z_{bc}^{a}=0,_{\shortmid }Z_{ab}^{i}=-\frac{g^{ij}}{2}\left[
~^{\circ }L_{aj}^{c}h_{cb}+~^{\circ }L_{bj}^{c}h_{ca}\right] ,  \notag
\end{eqnarray}%
for $\ \Xi _{jk}^{ih}=\frac{1}{2}(\delta _{j}^{i}\delta
_{k}^{h}-g_{jk}g^{ih}),~^{\pm }\Xi _{cd}^{ab}=\frac{1}{2}(\delta
_{c}^{a}\delta _{d}^{b}+h_{cd}h^{ab})$ and$~^{\circ
}L_{aj}^{c}=L_{aj}^{c}-e_{a}(N_{j}^{c}).$ If we work with nonholonomic
constraints on the dynamics/ geometry of gravity fields, it is more
convenient to use a N--adapted approach. For other purposes, it is preferred
to use only the Levi--Civita connection.

We compute also the curvature 2--form from (\ref{cart2}),%
\begin{eqnarray}
\widehat{\mathcal{R}}_{\ \gamma }^{\tau } &=&\widehat{\mathbf{R}}_{\ \gamma
\alpha \beta }^{\tau }\ \mathbf{e}^{\alpha }\wedge \ \mathbf{e}^{\beta }
\label{cform} \\
&=&\frac{1}{2}\widehat{R}_{\ jkh}^{i}e^{k}\wedge e^{h}+\widehat{P}_{\
jka}^{i}e^{k}\wedge \mathbf{e}^{a}+\frac{1}{2}\ \widehat{S}_{\ jcd}^{i}%
\mathbf{e}^{c}\wedge \mathbf{e}^{d},  \notag
\end{eqnarray}%
where the nontrivial N--adapted coefficients of curvature $\widehat{\mathbf{R%
}}_{\ \beta \gamma \tau }^{\alpha }$ of $\widehat{\mathbf{D}}$ are
\begin{eqnarray}
\widehat{R}_{\ hjk}^{i} &=&\mathbf{e}_{k}\widehat{L}_{\ hj}^{i}-\mathbf{e}%
_{j}\widehat{L}_{\ hk}^{i}+\widehat{L}_{\ hj}^{m}\widehat{L}_{\ mk}^{i}-%
\widehat{L}_{\ hk}^{m}\widehat{L}_{\ mj}^{i}-\widehat{C}_{\ ha}^{i}\
^{L}\Omega _{\ kj}^{a}  \label{cdcurv} \\
\widehat{P}_{\ jka}^{i} &=&e_{a}\widehat{L}_{\ jk}^{i}-\widehat{\mathbf{D}}%
_{k}\widehat{C}_{\ ja}^{i},\ \widehat{S}_{\ bcd}^{a}=e_{d}\widehat{C}_{\
bc}^{a}-e_{c}\widehat{C}_{\ bd}^{a}+\widehat{C}_{\ bc}^{e}\widehat{C}_{\
ed}^{a}-\widehat{C}_{\ bd}^{e}\widehat{C}_{\ ec}^{a}.  \notag
\end{eqnarray}%
Contracting the first and forth indices $\widehat{\mathbf{\mathbf{R}}}%
\mathbf{_{\ \beta \gamma }=}\widehat{\mathbf{\mathbf{R}}}\mathbf{_{\ \beta
\gamma \alpha }^{\alpha }}$, we get the N--adapted coefficients for the
Ricci tensor%
\begin{equation}
\widehat{\mathbf{\mathbf{R}}}\mathbf{_{\beta \gamma }=}\left( \widehat{R}%
_{ij},\widehat{R}_{ia},\widehat{R}_{ai},\widehat{R}_{ab}\right) .
\label{dricci}
\end{equation}%
The scalar curvature $\ ^{L}R=\widehat{R}$ of $\ \widehat{\mathbf{D}}$ is
\begin{equation}
\ ^{L}R=\ ^{L}\mathbf{g}^{\beta \gamma }\widehat{\mathbf{\mathbf{R}}}\mathbf{%
_{\beta \gamma }=\ \mathbf{g}^{\beta ^{\prime }\gamma ^{\prime }}\widehat{%
\mathbf{\mathbf{R}}}\mathbf{_{\beta ^{\prime }\gamma ^{\prime }}}.}
\label{scalc}
\end{equation}

The normal d--connection $\widehat{\mathbf{D}}$ (\ref{ndc}) defines a
canonical almost symplectic d--connection, $\widehat{\mathbf{D}}\equiv \
_{\theta }\widehat{\mathbf{D}},$ which is N--adapted to the effective
Lagrange and, related, almost symplectic structures, i.e. it preserves under
parallelism the splitting (\ref{whitney}), $_{\theta }\widehat{\mathbf{D}}_{%
\mathbf{X}}\ ^{L}\mathbf{\theta =}_{\theta }\widehat{\mathbf{D}}_{\mathbf{X}%
}\ \mathbf{\theta =}0$ and its torsion is constrained to satisfy the
conditions $\widehat{T}_{jk}^{i}=\widehat{T}_{bc}^{a}=0.$

\end{document}